\documentclass{amsproc}

\usepackage{amssymb, amsbsy, amsfonts,color}
\usepackage{amsmath}
\usepackage{natbib}

\allowdisplaybreaks[3]

\newcommand{\ev}{{\rm ev}}
\newcommand{\dd}{\mathfrak{d}}

\newcommand{\seqP}[1]{{S}(#1)}
\newcommand{\Shift}{{\mathcal S}}

\newcommand{\depth}[1]{\delta_{\KK}(#1)}
\newcommand{\depthQ}[1]{\delta_{\QQ}(#1)}
\newcommand{\DOS}{{\sf DOS}}
\newcommand{\SigmaP}{{\sf Sigma}}

\newcommand{\Sum}{\text{Sum}}
\newcommand{\Prod}{\text{Prod}}
\newcommand{\ProdSum}{\text{ProdSum}}

\newcommand{\MyFrame}[1]{
\noindent\fcolorbox[rgb]{0,0,0}{1,1,1}{
\begin{minipage}{12.2cm}#1\end{minipage}}
}

\newcommand{\RR}{\mathbb{A}}
\newcommand{\NN}{\mathbb{N}}
\newcommand{\ZZ}{\mathbb{Z}}
\newcommand{\QQ}{\mathbb{Q}}
\newcommand{\KK}{\mathbb{K}}
\newcommand{\FF}{\mathbb{F}}
\newcommand{\EE}{\mathbb{E}}
\newcommand{\GG}{\mathbb{G}}
\newcommand{\HH}{\mathbb{H}}
\newcommand{\DD}{\mathbb{D}}
\newcommand{\sigmaSE}{$\Sigma^*$}
\newcommand{\piE}{$\Pi$}
\newcommand{\pisiSE}{$\Pi\Sigma^*$}
\newcommand{\pisiDE}{$\Pi\Sigma^{\delta}$}
\newcommand{\sigmaDE}{$\Sigma^{\delta}$}
\newcommand{\vect}[1]{\boldsymbol{#1}}
\newcommand{\notion}[1]{{\it #1}}
\newcommand{\dfield}[2]{({#1},{#2})}
\newcommand{\const}[2]{{\rm const}_{#2}{#1}}
\newcommand{\seqR}{{S}(\KK)}
\newcommand{\seqQ}{{S}(\QQ)}
\newcommand{\set}[2]{{\{#1\,|\,#2\}}}

\newcommand{\fct}[3]{{#1:#2 \to #3}}

\newcounter{linectr}
\newenvironment{myEnumerate}{\begin{list}{(\arabic{linectr})}{\usecounter{linectr}
\labelwidth2.5ex\itemsep0ex\labelsep1ex\leftmargin3.ex\parskip0cm\topskip0cm\partopsep0cm
\listparindent0ex}}{\end{list}}

\newenvironment{myItemize}{\begin{list}{$\bullet$}{\usecounter{linectr}
\labelwidth2.5ex\itemsep0ex\labelsep1ex\leftmargin3.ex\parskip0cm\topskip0cm\partopsep0cm
\listparindent0ex}}{\end{list}}

\newtheorem{theorem}{Theorem}[section]
\newtheorem{lemma}[theorem]{Lemma}

\theoremstyle{definition}
\newtheorem{definition}[theorem]{Definition}
\newtheorem{example}[theorem]{Example}
\newtheorem{gencase}[theorem]{GENERAL CASE}

\theoremstyle{remark}
\newtheorem{remark}[theorem]{Remark}

\numberwithin{equation}{section}

\begin{document}

\title[Optimal Nested Sum Representations]{A Symbolic Summation Approach to Find\\
Optimal Nested Sum Representations}

\author{Carsten Schneider}
\address{Research Institute for Symbolic Computation\\
  Johannes Kepler University\\
  A-4040 Linz, Austria}
\email{Carsten.Schneider@risc.uni-linz.ac.at}
\thanks{Supported by the SFB-grant F1305 and the grant P20347-N18
of the Austrian FWF}


\begin{abstract}
We consider the following problem: Given a nested sum expression, find a sum representation such that the nested depth is minimal. We obtain a symbolic summation framework that solves this problem for sums defined, e.g., over hypergeometric, $q$-hypergeometric or mixed hypergeometric expressions. Recently, our methods have found applications in quantum field theory.
\end{abstract}


\maketitle

\section{Introduction}

Karr's algorithm~\cite{Karr:81,Karr:85} based on his difference field theory provides a general framework for symbolic summation. E.g., his algorithm, or a simplified version presented in~\cite{Schneider:05a}, covers summation over hypergeometric terms~\cite{Gosper:78,Zeilberger:91}, $q$-hypergeometric terms~\cite{PauleRiese:97} or mixed hypergeometric terms~\cite{Bauer:99}. More generally, indefinite nested product-sum expressions can be represented in his $\Pi\Sigma$-difference fields
which cover as special cases,
e.g., harmonic sums~\cite{Bluemlein:99,Vermaseren:99} or generalized nested harmonic sums~\cite{Moch:02}.


In this article much emphasize is put on the problem how these indefinite nested product-sum expressions can be simplified in a \pisiSE-field. E.g., with our algorithms we shall compute for the sum expression
\begin{align}\label{Equ:NestedSumExp}
A=\sum_{r=1}^n\frac{\displaystyle\sum_{l=1}^r\frac{H_l^2+H^{(2)}_l}{l}+\sum_{l=1}^r\frac{H_l}{l}}{r}
\end{align}
the alternative representation
\begin{equation}\label{Equ:BSol}
B=\tfrac{1}{12} \Big(H_n^4+2 H_n^3+6 (H_n+1) H^{(2)}_n H_n+3 \big({H^{(2)}_n}\big)^2+(8 H_n+4) H^{(3)}_n+6 H^{(4)}_n\Big)
\end{equation}
where $A(n)=B(n)$ for all $n\geq0$ and where the nested depth of $B$ is minimal; $H_n=\sum_{k=1}^n\frac{1}{k}$ denotes the $n$th harmonic numbers and $H^{(o)}_n=\sum_{k=1}^n\frac{1}{k^o}$ are the generalized harmonic numbers of order $o\geq1$.

In order to accomplish this task, we exploit a new difference field theory for symbolic summation~\cite{Schneider:08c} that refines Karr's $\Pi\Sigma$-fields to the so-called depth-optimal \pisiSE-fields.
In particular, we construct explicitly a difference ring monomorphism~\cite{Schneider:09b} which links elements from such a difference field to elements in the ring of sequences. Using this algorithmic machinery, we will derive for a given nested product-sum expression $A$ a nested product-sum expression $B$ with the following property: There is an explicit $\lambda\in\NN=\{0,1,2,\dots\}$ such that \begin{equation}\label{Equ:x=y}
A(k)=B(k)\quad \forall k\geq\lambda
\end{equation}
and among all such alternative representation for $A$ the depth of $B$ is minimal.

\medskip

From an applicational point of view our algorithms are able to produce  d'Alem\-bertian solutions~\cite{Noerlund,Abramov:94,Schneider:01a}, a subclass of Liouvillian
solutions~\cite{Singer:99}, of a given recurrence with minimal nested depth; for applications arising from particle physics see, e.g., \cite{Schneider:07h,Schneider:07i,Schneider:08e,Schneider:T09a,Schneider:09d}.
The presented algorithms are implemented in the summation package~\SigmaP~\cite{Schneider:07a}, that can be executed in the computer algebra system Mathematica.

\medskip

The general structure of this article is as follows. In Section~\ref{Sec:ProblemInSeq} we introduce the problem to find optimal sum representations which we supplement by concrete examples. In Section~\ref{Sec:SymbSumDF} we define depth-optimal \pisiSE-extensions and show how indefinite summation can be handled accordingly in such fields. After showing how generalized d'Alembertian extensions can be embedded in the ring of sequences in Section~\ref{Sec:Embed}, we are ready to prove that our algorithms produce sum representations with optimal nested depth in Section~\ref{Sec:OptimalRep}. Applications are presented in Section~\ref{Sec:dAlembert}.

\section{The problem description for indefinite nested sum expressions}\label{Sec:ProblemInSeq}

Inspired by~\cite{Buchberger:82,PauleNemes:97} one can consider the following general simplification problem. Let $X$ be a set of expressions (i.e., terms of certain types), let $\KK$ be a field\footnote{Subsequently, all fields and rings are commutative and contain the rational numbers $\QQ$.}, and let $\fct{\ev}{X\times\NN}{\KK}$ with $(x,n)\mapsto x(n)$ be a function. Here one considers the so called \notion{evaluation function} $\ev$ as a procedure that computes $x(n)$ for a given $x\in X$ and $n\in\NN$ in finite steps. In addition, we suppose that we are given a function $\fct{\dd}{X}{\NN}$ which measures the simplicity of the expressions in $X$; subsequently, we call such a triple $(X,\ev,\dd)$ also a \notion{(measured) sequence domain}; cf.~\cite{PauleNemes:97}.\\
In this setting the following problem can be stated: \textit{Given} $A\in X$; \textit{find} $B\in X$ and $\lambda\in\NN$ s.t.~\eqref{Equ:x=y} and s.t.\ among all such possible solutions $\dd(B)$ is minimal.

\medskip

In this article, the expressions $X$ are given in terms of indefinite nested sums and products and the measurement of simplicity is given by the nested depth of the occurring sum- and product-quantifiers. Subsequently, we shall make this more precise. Let $(X,\ev,\dd)$ be a measured sequence domain with $\fct{\ev}{X\times\NN}{\KK}$ and $\fct{\dd}{X}{\NN}$.

\begin{example}\label{Exp:RatCase}
Let $X=\KK(x)$ be a rational function field and define for $f=\frac{p}{q}\in\KK(x)$ with $p,q\in\KK[x]$ where $q\neq0$ and $p,q$ being coprime the evaluation function
\begin{equation}\label{Equ:EvalRat}
\ev(f,k)=\begin{cases}
0&\text{if }q(k)=0\\
\frac{p(k)}{q(k)}&\text{if }q(k)\neq0;
\end{cases}
\end{equation}
here $p(k),q(k)$ is the usual evaluation of polynomials at $k\in\NN$. In particular, we define $\dd(f)=1$ if $f\in\KK(x)\setminus\KK$ and $\dd(f)=0$ if $f\in\KK$. In the following $(\KK(x),\ev,\dd)$ is called the \notion{rational sequence domain}.
\end{example}

\begin{example}\label{Exp:MixedTerms}
Suppose that $\KK=\KK'(q_1,\dots,q_m)$ is a rational function field extension over $\KK'$ and consider the rational function field $X:=\KK(x,x_1,\dots,x_m)$ over $\KK$. Then for $f=\frac{p}{q}\in X$ with $p,q\in\KK[x,x_1,\dots,x_m]$ where $q\neq0$ and $p,q$ being coprime we define
\begin{equation}\label{Equ:EvalMixed}
\ev(f,k)=\begin{cases}
0&\text{if }q(k,q_1^k,\dots,q_m^k)=0\\
\frac{p(k,q_1^k,\dots,q_m^k)}{q(k,q_1^k,\dots,q_m^k)}&\text{if }q(k,q_1^k,\dots,q_m^k)\neq0.
\end{cases}
\end{equation}
Note that there is a $\delta\in\NN$ s.t.\ for all $k\in\NN$ with $k\geq\delta$ we have $q(k,q_1^k,\dots,q_m^k)\neq0$; for an algorithm that determines $\delta$ see~\cite[Sec.~3.7]{Bauer:99}.
In particular, we define $\dd(f)=0$ if $f\in\KK$, and $\dd(f)=1$ if $f\notin\KK$. In the following $(\KK(x,x_1,\dots,x_m),\ev,\dd)$ is called \notion{$q$-mixed sequence domain}. Note: If $m=0$, we are back to the rational sequence domain. If we restrict to the setting $\KK(x_1,\dots,x_m)$ which is free of $x$, it is called \notion{$q$-rational sequence domain}.
\end{example}

More generally, $X$ can contain hypergeometric, $q$-hypergeometric or mixed hypergeometric terms; for instance see Example~\ref{Exp:Embedding4}.
Over such a set $X$ we consider the set of \notion{(indefinite nested) product-sum expressions} denoted by $\ProdSum(X)$ which is defined as follows. Let $\oplus$ ,$\otimes$, $\Sum$ ,$\Prod$ be operations with the signatures
$$\begin{array}[t]{llcl}
\oplus: &\ProdSum(X)\times\ProdSum(X)&\rightarrow&\ProdSum(X)\\
\otimes: &\ProdSum(X)\times\ProdSum(X)&\rightarrow&\ProdSum(X)\\
\Sum: &\NN\times\ProdSum(X)&\rightarrow&\ProdSum(X)\\
\Prod: &\NN\times\ProdSum(X)&\rightarrow&\ProdSum(X).
\end{array}$$
Then $\ProdSum(X)\supseteq X$ is the smallest set that satisfies the following rules:
\begin{myEnumerate}
\item For any $f,g\in\ProdSum(X)$, $f\oplus g\in\ProdSum(X)$ and $f\otimes g\in\ProdSum(X)$.
\item For any $f\in\ProdSum(X)$ and any $r\in\NN$, $\Sum(r,f)\in\ProdSum(X)$ and $\Prod(r,f)\in\ProdSum(X)$.
\end{myEnumerate}

The set of all expressions in $\Prod\Sum(X)$ which are free of  $\Prod$ is denoted by $\Sum(X)$. $\Sum(X)$ is called the set of \notion{(indefinite nested) sum expressions over $X$}.

\begin{example}\label{Exp:RatExpression}
Given $(X,\ev,\dd)$ from Example~\ref{Exp:RatCase} with $\KK=\QQ$ and $X=\QQ(x)$ the following indefinite nested sum expressions are in $\Sum(\QQ(x))$:
\begin{align*}
E_1&=\frac{1}{x},\quad E_2=\Sum(1,\frac{1}{x}),\text{ and}\\
A&=\Sum(1,\frac{1}{x}\Big(\Sum(1,\frac{1}{x}\big(\Sum(1,\frac{1}{x})^2\oplus \Sum(1,\frac{1}{x^2})\big))\oplus \Sum(1,\Sum(1,\frac{1}{x}))\Big)).
\end{align*}
\end{example}

\noindent Finally, $\ev$ and $\dd$ are extended from $X$ to $\fct{\ev'}{\ProdSum(X)\times\NN}{\ProdSum(X)}$ with $(x,n)\mapsto x(n)$ and $\fct{\dd'}{\ProdSum(X)}{\NN}$ as follows.
\begin{myEnumerate}
\item For $f\in X$ we set $\dd'(f):=\dd(f)$ and $\ev'(f,k):=\ev(f,k)$.
\item For $f,g\in\ProdSum(X)$ we set $\dd'(f\oplus g)=\dd'(f\otimes g):=\max(\dd'(f),\dd'(g)$,
$$\ev'(f\oplus g,k):=\ev'(f,k)+\ev'(g,k)\quad\text{and}\quad\ev'(f\otimes g,k):=\ev'(f,k)\,\ev'(g,k);$$
here the operations on the right hand side are from the field $\KK$.
\item For $r\in\NN$, $f\in\ProdSum(X)$ define $\dd'(\Sum(r,f))=\dd'(\Prod(r,f)):=\dd'(f)+1$,
\footnotetext{Note that $\ev'(\Prod(r,f),k)$ might be $0$ if $r$ is too small. Later, products will be used only as described in Ex.~\ref{Exp:Embedding4} or in the General case~\ref{Exp:GeneralSequenceDom}; there we will take care of the bound $r$ by~\eqref{Equ:ChooseLowerBound}.}
\begin{align*}
\ev'(\Sum(r,f),k)&=\sum_{i=r}^k\ev'(f,i)&\text{and\footnotemark}&&
\ev'(\Prod(r,f),k)&=\prod_{i=r}^k\ev'(f,i).
\end{align*}
\end{myEnumerate}
Since $\ev'$ and $\ev$, resp.\ $\dd$ and $\dd'$, agree on $X$, we do not distinguish them anymore. Subsequently, $(\ProdSum(X),\ev,\dd)$ (resp.~$(\Sum(X),\ev,\dd)$) is called the \notion{product-sum sequence domain over $X$} (resp.\ \notion{sum sequence domain over $X$}).

\begin{example}\label{Exp:RatExpressionUsual}
The expressions from Example~\ref{Exp:RatExpression} are evaluated as
\begin{align*}
\ev(E_1,k)&=E_1(k)=\frac{1}{k},\quad \ev(E_2,k)=E_2(k)=\sum_{i=1}^k\ev(\frac{1}{x},i)=\sum_{i=1}^k\frac{1}{i},\quad\text{ and}\\ \ev(A,k)&=A(k)=\sum_{r=1}^k\frac{\displaystyle\sum_{l=1}^r\frac{\Big(\sum_{i=1}^l\frac{1}{i}\Big)^2+\sum_{i=1}^l\frac{1}{i^2}}{l}+\sum_{l=1}^r\frac{\sum_{i=1}^l\frac{1}{i}}{l}}{r}.
\end{align*}
We have $\dd(E_1)=1$, $\dd(E_2)=2$ and $\dd(A)=4$.
\end{example}

\noindent Usually, we stick to the following more convenient and frequently used notation.

\begin{myItemize}
\item We write, e.g.,
$E=a\oplus \Sum(1,c\,\Sum(2,b))\in\Sum(X)$
with $a,b,c\in X$ in the form
$$E'=\ev(a,n)\oplus\sum_{i=1}^n\ev(c,i)\sum_{j=2}^i\ev(b,j)$$
for a {\sf symbolic variable $n$}. Clearly, fixing the variable $n$, the two encodings $E$ and $E'$ can be transformed into each other; if we want to emphasize the dependence on $n$, we also write $E'\in\Sum_n(X)$.
\item Even more, by abuse of notation, we use instead of $\oplus$ and $\otimes$ the usual field operations in $\KK$. This ``sloppy'' notation immediately produces the evaluation mechanism: $\ev(E,k)=E(k)$ for a concrete integer $k\in\NN$ is produced by substitution in $E'$ the variable $n$ with the concrete value $k\in\NN$.

\item Finally, whenever possible, the evaluation $\ev(a,n)$ for some $a\in X$ is expressed by well known functions, like, e.g., $\ev(1/(x+1),n)=\frac{1}{(n+1)}$ (Ex.~\ref{Exp:RatCase}) or $\ev(x_i,n)=q_i^n$ for $1\leq i\leq e$ (Ex.~\ref{Exp:MixedTerms}).
\end{myItemize}

\begin{example}\label{Exp:ASum}
We write $E_1,E_2,A\in\Sigma_n(\QQ(x))$ from Example~\ref{Exp:RatExpressionUsual} in the more convenient notation
$E_1=\frac{1}{n}$, $E_2=H_n$ and~\eqref{Equ:NestedSumExp}; note that for $E_1$ we must require that $n$ is only evaluated for $n\geq1$.
\end{example}

Let $(X,\ev,\dd)$ be a measured sequence domain and consider the sum sequence domain $(\Sum(X),\ev,\dd)$ over $X$.
We define the \notion{$\Sum(X)$-optimal depth} of $A\in\Sum(X)$ as
$$\min\{\dd(B)|B\in\Sum(X)\text{ s.t.~\eqref{Equ:x=y} for some } \lambda\in\NN\}.$$
Then we are interested in the following problem.

\medskip

\noindent\MyFrame{\noindent\textsf{\DOS: Depth Optimal Simplification.}
\noindent\textit{Given} $A\in\Sum(X)$; \textit{find} $B\in\Sum(X)$ and $\lambda\in\NN$ s.t.~\eqref{Equ:x=y} and s.t.\ $\dd(B)$ is the $\Sum(X)$-optimal depth of $A$.}


\begin{example}\label{Equ:BExpression}
Consider, e.g., $A\in\Sum(\QQ(x))$ from Example~\ref{Exp:ASum}. Then with our algorithms, see Example~\ref{Exp:GoodConstruction}, we find
$B\in\Sum(\KK(x))$ with~\eqref{Equ:BSol} such that $A(n)=B(n)$ for all $n\in\NN$. At this point it is easy to see that $B$ cannot be expressed with depth$\leq1$, and thus $B$ is a solution of~\DOS. In Section~\ref{Sec:OptimalRep} we will show that this fact is an immediate consequence of our algebraic construction. Summarizing, $2$ is the $\Sum(\QQ(x))$-optimal depth of $A$.
\end{example}

\medskip

We shall solve problem~\DOS\ algorithmically, if $X$ is, e.g., the rational sequence domain (Ex.~\ref{Exp:RatCase}) or the $q$-mixed sequence domain (Ex.~\ref{Exp:MixedTerms}). More generally, $X$ might be a sequence domain in which a finite number of objects from $\ProdSum(X')$ can be represented; in this setting $X'$ might be the rational, $q$-rational or $q$-mixed sequence domain. Note that the General~case~\ref{Exp:GeneralSequenceDom} (page~\pageref{Exp:GeneralSequenceDom}) includes most of the ($q$--)hypergeometric or $q$-mixed hypergeometric terms (see Ex.~\ref{Exp:Embedding4}).

\section{Step I: Reducing the problem to difference fields by telescoping}\label{Sec:SymbSumDF}

Let $(X,\ev,\dd)$ be a measured sequence domain. Then for $f\in\ProdSum(X)$ and $r\in\NN$ the sum  $S=\sum_{k=r}^nf(k)\in\ProdSum_n(X)$ satisfies the recurrence relation
\begin{equation}\label{Equ:SumRel}
S(n+1)=S(n)+f(n+1)\quad\forall n\geq r,
\end{equation}
and the product $P=\sum_{k=r}^nf(k)\in\ProdSum_n(X)$ satisfies the recurrence relation
\begin{equation}\label{Equ:ProdRel}
P(n+1)=f(n+1)P(n)\quad\forall n\geq r.
\end{equation}
As a consequence, we can define a \notion{shift operator} acting on the expressions $S(n)$ and $P(n)$. Subsequently, we shall restrict to sequence domains $X$ such that the sums and products $S(n)$ and $P(N)$ can be modeled in difference rings.

\smallskip


In general, a \notion{difference ring} (resp.~\notion{difference field})
$\dfield{\RR}{\sigma}$ is defined as a ring $\RR$ (resp.~field) with a ring
automorphism (resp.~field automorphism) $\fct{\sigma}{\RR}{\RR}$.
The set of constants
$\const{\RR}{\sigma}=\set{k\in\RR}{\sigma(k)=k}$ forms a subring\footnote{Subsequently, we assume that
$\const{\RR}{\sigma}$ is always a field, which we usually denote by $\KK$. Note that this implies that $\QQ$ is a subfield of $\KK$.} (resp.\ subfield) of $\RR$.
We call $\const{\RR}{\sigma}$ the \notion{constant field} of
$\dfield{\RR}{\sigma}$. A difference ring (resp.~difference field) $\dfield{\EE}{\sigma}$ is
a \notion{difference ring extension} (resp.\ \notion{difference field
extension}) of a difference ring (resp.\ difference field)
$\dfield{\RR}{\sigma'}$ if $\RR$ is a subring (resp.~subfield) of
$\EE$ and $\sigma'(f)=\sigma(f)$ for all $f\in\RR$;
we call $\dfield{\RR}{\sigma'}$ also a sub-difference ring (resp. field) of $\dfield{\EE}{\sigma}$.  Since $\sigma$
and $\sigma'$ agree on $\RR$, we do not distinguish them anymore.

\begin{example}\label{Exp:RatDF}
For the rational function field $\KK(x)$ we can define uniquely the  automorphism $\fct{\sigma}{\KK(x)}{\KK(x)}$ s.t.\ $\sigma(x)=x+1$ and s.t.\ $\sigma(c)=c$ for all $c\in\KK$; $\dfield{\KK(x)}{\sigma}$ is called the \notion{rational difference field over $\KK$}.
\end{example}

\begin{example}\label{Exp:MixedDF}
For the rational function field $\FF:=\KK(x,x_1,\dots,x_m)$ from Ex.~\ref{Exp:MixedTerms} we can define uniquely the field automorphism $\fct{\sigma}{\FF}{\FF}$ such that $\sigma(x)=x+1$ and $\sigma(x_i)=q_i\,x_i$ for all $1\leq i\leq m$ and such that $\sigma(c)=c$ for all $c\in\KK$. The difference field $\dfield{\FF}{\sigma}$ is also called the \notion{$q$-mixed difference field over $\KK$}.
\end{example}

\noindent Then any expression in $\ProdSum(\KK(x))$ (resp.\ in \ProdSum( $\KK(x,x_1,\dots,x_m))$) with its shift behavior can be modeled by defining a tower of difference field extensions over $\dfield{\KK(x)}{\sigma}$ (resp. of $\dfield{\KK(x,x_1,\dots,x_m)}{\sigma}$). Subsequently, we restrict to those extensions in which the constants remain unchanged. We confine to
\pisiSE-extensions~\cite{Schneider:01a} being slightly less general but covering all sums and products treated explicitly in Karr's $\Pi\Sigma$-extensions~\cite{Karr:85}.

\begin{definition}
A difference field extension $\dfield{\FF(t)}{\sigma}$ of
$\dfield{\FF}{\sigma}$ is called a \notion{\pisiSE-extension} if
both difference fields share the same field of constants, $t$~is
transcendental over~$\FF$, and $\sigma(t)=t+a$ for some $a\in\FF^*$
or $\sigma(t)=a\, t$ for some $a\in\FF^*$. If
$\sigma(t)/t\in\FF$ (resp.\ $\sigma(t)-t\in\FF$), we call the
extension also a \notion{\piE-extension}
(resp.~\notion{\sigmaSE-extension}). In short, we say that
$\dfield{\FF(t_1)\dots(t_e)}{\sigma}$ is a \notion\pisiSE-extension
(resp.~\piE-extension, \sigmaSE-extension) of $\dfield{\FF}{\sigma}$
if the extension is given by a tower of \pisiSE-extensions
(resp.~\piE-extensions, \sigmaSE-extensions).
We call
a \pisiSE-extension $\dfield{\FF(t_1)\dots(t_e)}{\sigma}$ of
$\dfield{\FF}{\sigma}$ with $\sigma(t_i)=\alpha_i\,t_i+\beta_i$ \notion{generalized d'Alembertian}, or in short \notion{polynomial}, if
$\alpha_i\in\FF^*$ and $\beta_i\in\FF[t_1,\dots,t_{i-1}]$ for all
$1\leq i\leq e$.
A
\notion{\pisiSE-field} $\dfield{\KK(t_1)\dots(t_e)}{\sigma}$ over
$\KK$ is a \pisiSE-extension of $\dfield{\KK}{\sigma}$ with constant
field $\KK$.
\end{definition}

\begin{remark}
If $\dfield{\FF(t_1)\dots(t_e)}{\sigma}$ is a polynomial \pisiSE-extension of $\dfield{\FF}{\sigma}$ then it follows that $\dfield{\FF[t_1]\dots[t_e]}{\sigma}$ is a difference ring extension of $\dfield{\FF}{\sigma}$.
\end{remark}

\smallskip

\textbf{Karr's approach.} The following result from~\cite{Karr:81} tells us how one can design a \pisiSE-field for a given product-sum expression.

\begin{theorem}\label{Thm:PiSigma}
Let $\dfield{\FF(t)}{\sigma}$ be a difference field extension of
$\dfield{\FF}{\sigma}$ with $\sigma(t)=a\,t+f$ where $a\in\FF^*$ and $f\in\FF$. Then the following holds.
\begin{myEnumerate}
\item $\dfield{\FF(t)}{\sigma}$ is a \sigmaSE-extension of $\dfield{\FF}{\sigma}$ iff $a=1$ and there is no
$g\in\FF$ s.t.
\begin{equation}\label{Equ:Telef}
\sigma(g)=g+f.
\end{equation}
\item $\dfield{\FF(t)}{\sigma}$ is a \piE-extension of $\dfield{\FF}{\sigma}$ iff $t\neq0$, $f=0$ and there are no $g\in\FF^*$ and $m>0$
such that $\sigma(g)=a^m g$.
\end{myEnumerate}
\end{theorem}

E.g., with Theorem~\ref{Thm:PiSigma} it is easy to see that the difference fields from Examples~\ref{Exp:RatDF}
and~\ref{Exp:MixedDF} are \pisiSE-fields over $\KK$.

From the algorithmic point of view we emphasize the following: For a given \pisiSE-field $\dfield{\FF}{\sigma}$ and $f\in\FF$, Karr's summation algorithm~\cite{Karr:81} can compute a solution $g\in\FF$ for the telescoping equation~\eqref{Equ:Telef}, or it outputs that such a solution in $\FF$ does not exist; for a simplified version see~\cite{Schneider:05a}. In this case, we can adjoin a new \sigmaSE-extension which produces by construction a solution for~\eqref{Equ:Telef}.\\
Summarizing, Karr's algorithm in combination with Theorem~\ref{Thm:PiSigma} enables one to construct algorithmically a \pisiSE-field that encodes the shift behavior of a given indefinite nested sum expression.

\begin{example}\label{Exp:BadConstruction}
We start with the \pisiSE-field $\dfield{\QQ(x)}{\sigma}$ over $\QQ$ with $\sigma(x)=x+1$. Now we consider the sum expressions of $A$ in~\eqref{Equ:NestedSumExp}, say in the order
\begin{equation}\label{Equ:SumOrder}
\begin{split}
\stackrel{(1)}{\to}H_n=\sum_{i=1}^n\frac{1}{i}
\stackrel{(2)}{\to}S:=\sum_{i=1}^n\frac{H_i}{i}
\stackrel{(3)}{\to}H^{(2)}_n=\sum_{i=1}^n\frac{1}{i^2}
\stackrel{(4)}{\to}T:=\sum_{i=1}^n\frac{H_i^2+H^{(2)}_i}{i}
\stackrel{(5)}{\to}A,
\end{split}
\end{equation}
and represent them in terms of \sigmaSE-extensions following Theorem~\ref{Thm:PiSigma}.1.
\begin{myEnumerate}
\item Using, e.g., Gosper's algorithm~\cite{Gosper:78}, Karr's algorithm~\cite{Karr:81} or a simplified version of it presented in~\cite{Schneider:05a}, we check that there is no $g\in\QQ(x)$ with $\sigma(g)=g+\frac{1}{x+1}$. Hence, by Theorem~\ref{Thm:PiSigma}.1 we adjoin $H_n$ in form of the \sigmaSE-extension $\dfield{\QQ(x)(h)}{\sigma}$ of $\dfield{\QQ(x)}{\sigma}$ with $\sigma(h)=h+\frac{1}{x+1}$; note that the shift behavior $H_{n+1}=H_n+\frac{1}{n+1}$ is reflected by the automorphism $\sigma$.
\item With the algorithms from~\cite{Karr:81} or~\cite{Schneider:05a} we show that there is no $g\in\QQ(x)(h)$ with $\sigma(g)=g+\frac{\sigma(h)}{x+1}$. Thus we take the \sigmaSE-extension $\dfield{\QQ(x)(h)(s)}{\sigma}$ of $\dfield{\QQ(x)(h)}{\sigma}$ with $\sigma(s)=s+\frac{\sigma(h)}{x+1}$ and express $S$ by $s$.
\item With the algorithms from above, we find $g=2s-h^2\in\QQ(x)(h)(s)$ with $\sigma(g)=g+\frac{1}{(x+1)^2}$, and represent\footnote{Note that there is no way to adjoin a \sigmaSE-extension $h_2$ of the desired type $\sigma(h_2)=h_2+1/(x+1)^2$, since otherwise $\sigma(g-h_2)=(g-h_2)$, i.e., $\const{\QQ(x)(h)(s)(h_2)}{\sigma}\neq\QQ$.} $H^{(2)}_n$ by $g$.
\item There is no $g\in\QQ(x)(h)(s)$ with $\sigma(g)=g+2\frac{\sigma(s)}{x+1}$; thus we rephrase $T$ as $t$ in the \sigmaSE-extension $\dfield{\QQ(x)(h)(s)(t)}{\sigma}$ of $\dfield{\QQ(x)(h)(s)}{\sigma}$ with $\sigma(t)=t+2\frac{\sigma(s)}{x+1}$.
\item There is no $g\in\QQ(x)(h)(s)(t)$ s.t.\ $\sigma(g)=g+\frac{\sigma(s+t)}{x+1}$; thus we represent $A$ with $a$ in the \sigmaSE-ext.\ $\dfield{\QQ(x)(h)(s)(t)(a)}{\sigma}$ of $\dfield{\QQ(x)(h)(s)(t)}{\sigma}$ with $\sigma(a)=a+\frac{\sigma(s+t)}{x+1}$.
\end{myEnumerate}
Reformulating $a$ as a sum expression (for more details see Section~\ref{Sec:Embed}) yields
\begin{equation}\label{Equ:WResult}
W=\sum_{r=1}^n\frac{\displaystyle\sum_{l=1}^r\frac{\displaystyle2\sum_{i=1}^l\frac{H_i}{i}}{l}+\sum_{l=1}^r\frac{H_l}{l}}{r}
\end{equation}
with $A(n)=W(n)$ for all $n\in\NN$.
\end{example}

We remark that the sums occurring in $W$ pop up only in the numerator. Here the following result plays an important role.

\begin{theorem}[\cite{Schneider:09b},Thm.~2.7]\label{Thm:PolyClosure}
Let $\dfield{\FF(t_1)\dots(t_e)}{\sigma}$ be a polynomial \pisiSE-ex\-tension
of $\dfield{\FF}{\sigma}$; let $\RR=\FF[t_1,\dots,t_e]$. Then for all $g\in\RR$,
$\sigma(g)-g\in\RR$ iff
$g\in\RR$.
\end{theorem}

\noindent Namely, if, e.g., $A$ consists only of sums that occur in the numerator, then by solving iteratively the telescoping problem, it is guaranteed that also the telescoping solutions will have only sums that occur in the numerators.

\begin{remark}
Similar to the sum case, there exit algorithms~\cite{Karr:81} which can handle the product case; for details and technical problems we refer to~\cite{Schneider:05c}. Note that \piE-extensions will occur later only in the frame of General case~\ref{Exp:GeneralSequenceDom}. At this point one has explicit control how the sequence domain $(X,\ev,\dd)$ for $\Sum(X)$ is defined.
\end{remark}

\medskip

\textbf{A depth-refined approach.} The depth of $W$ in~\eqref{Equ:WResult} is reflected by the nested depth of the underlying difference field constructed in Example~\ref{Exp:BadConstruction}.

\begin{definition}\label{Def:DepthDF} Let $\dfield{\FF}{\sigma}$ be a \pisiSE-field over $\KK$
with $\FF:=\KK(t_1)\dots(t_e)$ where $\sigma(t_i)=a_i\,t_i$ or
$\sigma(t_i)=t_i+a_i$ for $1\leq i\leq e$. The
\notion{depth function for elements of $\FF$},
$\fct{\delta_{\KK}}{\FF}{\NN}$, is defined as follows.
\begin{myEnumerate}
\item For any $g\in\KK$, $\depth{g}:=0$.

\item If $\delta_{\KK}$ is defined for
$\dfield{\KK(t_1)\dots(t_{i-1})}{\sigma}$ with $i>1$, we define
$\depth{t_{i}}:=\depth{a_i}+1$; for
$g=\frac{g_1}{g_2}\in\KK(t_1)\dots(t_i)$, with $g_1,g_2\in\KK[t_1,\dots,t_{i}]$ coprime, we define
$$\depth{g}:=
\max(\{\depth{t_j}|1\leq j\leq i\text{ and }t_j\text{ occurs in
}g_1\text{ or }g_2\}\cup\{0\}).$$
\end{myEnumerate}
The \notion{extension depth} of a \pisiSE-extension $\dfield{\FF(x_1)\dots(x_r)}{\sigma}$ of $\dfield{\FF}{\sigma}$ is defined by $\max(\depth{x_1},\dots,\depth{x_r},0).$
\end{definition}

\begin{example}
In Example~\ref{Exp:BadConstruction} we have $\depthQ{x}=1$, $\depthQ{h}=2$, $\depthQ{s}=3$, $\depthQ{t}=4$, and $\depthQ{a}=5$.
\end{example}

With the approach sketched in Example~\ref{Exp:BadConstruction} we obtain an alternative sum representation $W(n)$ for $A(n)$ with larger depth.
Motivated by such problematic situations, Karr's \pisiSE-fields have been refined in the following way; see~\cite{Schneider:05f,Schneider:08c}.

\begin{definition}
Let $\dfield{\FF}{\sigma}$ be a \pisiSE-field over $\KK$. A difference field extension $\dfield{\FF(s)}{\sigma}$ of
$\dfield{\FF}{\sigma}$ with $\sigma(s)=s+f$ is called
\notion{depth-optimal} \sigmaSE-extension, in short
\sigmaDE-extension, if there is no \sigmaSE-extension $\dfield{\EE}{\sigma}$ of $\dfield{\FF}{\sigma}$ with extension depth $\leq\depth{f}$ such that there is a $g\in\EE$ as in~\eqref{Equ:Telef}. A \pisiSE-extension
$\dfield{\FF(t_1)\dots(t_e)}{\sigma}$ of $\dfield{\FF}{\sigma}$ is depth-optimal, in short a \pisiDE-extension, if all \sigmaSE-extensions are
depth-optimal. A \pisiDE-field consists of \piE- and \sigmaDE-extensions.
\end{definition}

\noindent Note that a \sigmaDE-extension is a \sigmaSE-extension by Theorem~\ref{Thm:PiSigma}.1. Moreover, a \pisiSE-field $\dfield{\FF}{\sigma}$ with depth $\leq2$ and $x\in\FF$ such that $\sigma(x)=x+1$ is always depth-optimal; see~\cite[Prop.~19]{Schneider:08c}. \textit{In particular, the rational and the $q$-mixed difference fields from the Examples~\ref{Exp:RatDF}
and~\ref{Exp:MixedDF} are \pisiDE-fields over $\KK$.}
\smallskip

\noindent Given any \pisiDE-field, we obtain the following crucial property which will be essential to solve problem~DOS.

\begin{theorem}[\cite{Schneider:08c},Result~3]\label{Thm:DepthStable}
Let $\dfield{\FF}{\sigma}$ be a \pisiDE-field over $\KK$.
Then for any $f,g\in\FF$ such that~\eqref{Equ:Telef}
we have
\begin{equation}\label{Equ:DepthStable}
\depth{f}\leq\depth{g}\leq\depth{f}+1.
\end{equation}
\end{theorem}

\noindent In other words, in a given \pisiDE-field we can guarantee that the depth of a telescoping solution is not bigger than the depth of the sum itself.

\begin{example}\label{Exp:GoodConstruction}
We consider again the sum expressions in~\eqref{Equ:SumOrder}, but this time we use the refined algorithm presented in~\cite{Schneider:08c}.
\begin{myEnumerate}
\item As in Example~\ref{Exp:GoodConstruction} we compute the \pisiDE-field $\dfield{\QQ(x)(h)}{\sigma}$ and  represent $H_n$ with $h$. From this point on, our new algorithm works differently.

\item Given $\dfield{\QQ(x)(h)}{\sigma}$, we find the \sigmaDE-extension $\dfield{\QQ(x)(h)(h_2)}{\sigma}$ of $\dfield{\QQ(x)(h)}{\sigma}$ with $\sigma(h_2)=h_2+\frac{1}{(x+1)^2}$ in which we find $s'=\frac{1}{2}(h^2+h_2)$ such that $\sigma(s')-s'=\frac{\sigma(h)}{x+1}$. Hence we represent $S$ by $s'$.
\item $H^{(2)}_n$ can be represented by $h_2$ in the already constructed \pisiDE-field.
\item Our algorithm finds the \sigmaDE-exten\-sion $\dfield{\QQ(x)(h)(h_2)(h_3)}{\sigma}$ of $\dfield{\QQ(x)(h)(h_2)}{\sigma}$ with $\sigma(h_3)=h_3+\frac{1}{(x+1)^3}$ together with $t'=\frac{1}{3}(h^3+3 hh_2+2h_3)$ such that $\sigma(t')-t'=\frac{\sigma(h^2+h_2)}{x+1}$; hence we rephrase $T$ as $t'$.
\item Finally, we find the \sigmaDE-ext.\ $\dfield{\QQ(x)(h)(h_2)(h_3)(h_4)}{\sigma}$ of $\dfield{\QQ(x)(h)(h_2)(h_3)}{\sigma}$ with $\sigma(h_4)=h_4+\frac{1}{(x+1)^4}$ and get $a'=
\tfrac{1}{12}(h^4+2h^3+6(h+1)h_2h+3h_2^2+(8h+4)h_3+6h_4)$ s.t.\ $\sigma(a')-a'=\frac{\sigma(t'+s')}{x+1}$; $A$ is represented by $a'$.
\end{myEnumerate}
Reinterpreting $a'$ as a sum expression gives $B$ in~\eqref{Equ:BSol}; see also Example~\ref{Equ:BExpression}.
\end{example}

\noindent To sum up, we can compute step by step a \pisiDE-field in which we can represent nested sum expressions. To be more precise, we will exploit the following

\begin{theorem}[\cite{Schneider:08c},Result~1]\label{Thm:ConstructDeltaSigma}
Let $\dfield{\FF}{\sigma}$ be a \pisiDE-field over $\KK$ and $f\in\FF$.
\begin{myEnumerate}
\item There is a
\sigmaDE-extension $\dfield{\EE}{\sigma}$ of $\dfield{\FF}{\sigma}$ in which we have $g\in\EE$ such that~\eqref{Equ:Telef}; $\dfield{\EE}{\sigma}$ and $g$ can be given explicitly if $\KK$ has the form as stated in Remark~\ref{Remark:KAssum}.

\item Suppose that  $\dfield{\FF}{\sigma}$ with $\FF=\GG(y_1,\dots,y_r)$ is a polynomial \pisiDE-extension of $\dfield{\GG}{\sigma}$. If $f\in\GG[y_1\,\dots,y_r]$, then $\dfield{\EE}{\sigma}$ from part~(1) can be given as a polynomial \pisiDE-ex\-tension of $\dfield{\GG}{\sigma}$; if $\EE=\FF(t_1,\dots,t_e)$, then  $g\in\GG[y_1,\dots,y_r][t_1,\dots,t_e]$.
\end{myEnumerate}
\end{theorem}

\begin{remark}\label{Remark:KAssum}
From the computational point of view certain operations must be carried out in the constant field\footnote{Actually, we require the same computational properties as for Karr's summation algorithm; see~\cite[Thm.~9]{Karr:81} or~\cite[Thm.~3]{Schneider:06d}.}. For instance, Theorem~\ref{Thm:ConstructDeltaSigma} is completely constructive, if $\KK$ is of the following from: $\KK=\RR(q_1,\dots,q_m)$ is a rational function field with variables $q_1,\dots,q_m$ over an algebraic number field $\RR$. Due to the restrictions of the computer algebra system Mathematica, the implementation in \SigmaP~\cite{Schneider:07a} works only optimal if $\RR=\QQ$, i.e., $\KK=\QQ(q_1,\dots,q_m)$.
\end{remark}

\section{Step II: Reinterpretation as product-sum expressions}\label{Sec:Embed}

Let $(X,\ev,\dd)$, e.g., be the $q$-mixed sequence domain from Example~\ref{Exp:MixedTerms} with $X=\KK(x,x_1,\dots,x_m)$ and let $\dfield{\FF}{\sigma}$ with $X=\FF$ be the $q$-mixed difference field from Example~\ref{Exp:MixedDF}. Moreover, take $A\in\Sum(X)$.

Then in the previous section we have demonstrated how one can compute a polynomial \sigmaDE-extension $\dfield{\FF(t_1)\dots(t_e)}{\sigma}$ of $\dfield{\FF}{\sigma}$ in which one can model the shift-behavior of $A$ by an element $a\in\FF[t_1,\dots,t_e]$. Then, as illustrated in Example~\ref{Exp:GoodConstruction}, we were able to reinterpret $a$ as an element from $B\in\Sum(X)$ such that $\depth{a}=\dd(B)$ and such that~\eqref{Equ:x=y} where $\lambda\in\NN$ could be given explicitly.

In order to accomplish this task algorithmically, we will supplement the construction of the difference field $\dfield{\FF(t_1)\dots(t_e)}{\sigma}$ by defining in addition an explicitly given difference ring monomorphism.
Namely, following~\cite{Schneider:09b} we will embed the difference ring
$\dfield{\FF[t_1,\dots,t_e]}{\sigma}$ into the ring of sequences by a so-called $\KK$-monomorphism.
It turns out that any element $h\in\FF[t_1,\dots,t_e]$ can be mapped injectively to $\langle\ev(H,k)\rangle_{k\geq0}$ for some properly chosen expression $H\in\ProdSum(X)$.

\medskip
Subsequently, we define the ring of sequences and $\KK$-monomorphisms.
Let $\KK$ be a field and consider the set of sequences
$\KK^{\NN}$ with elements $\langle a_n\rangle_{n\geq0}=\langle
a_0,a_1,a_2,\dots\rangle$, $a_i\in\KK$. With component-wise addition and multiplication we obtain a commutative ring; the field $\KK$ can be
naturally embedded by identifying $k\in\KK$ with the sequence
$\langle k,k,k,\dots\rangle$; we write $\vect{0}=\langle0,0,0,\dots\rangle$.

\noindent We follow the construction
from~\cite[Sec.~8.2]{AequalB} in order to turn the
shift
\begin{equation}\label{Equ:ShiftOp}
\Shift:{\langle a_0,a_1,a_2,\dots\rangle}\mapsto{\langle
a_1,a_2,a_3,\dots\rangle}
\end{equation}
to an automorphism: We define an equivalence relation
$\sim$ on $\KK^{\NN}$ with $\langle a_n\rangle_{n\geq0}\sim
\langle b_n\rangle_{n\geq0}$ if there exists a $d\geq 0$ such that
$a_k=b_k$ for all $k\geq d$. The equivalence classes form a ring
which is denoted by $\seqR$; the elements of $\seqR$ (also called germs) will be
denoted, as above, by sequence notation.
Now it is immediate that
$\fct{\Shift}{\seqR}{\seqR}$ with~\eqref{Equ:ShiftOp} forms a ring automorphism. The difference ring $\dfield{\seqR}{\Shift}$ is called the \notion{ring of sequences (over $\KK$)}.

\medskip

A \notion{difference ring homomorphism} $\fct{\tau}{\RR_1}{\RR_2}$ between difference
rings $\dfield{\RR_1}{\sigma_1}$ and $\dfield{\RR_2}{\sigma_2}$ is a
ring homomorphism such that $\tau(\sigma_1(f))=\sigma_2(\tau(f))$ for all
$f\in\RR_1$. If $\tau$ is injective, we call $\tau$ a \notion{difference ring monomorphism}.

Let $\dfield{\RR}{\sigma}$ be a difference ring with constant field
$\KK$. Then a difference ring homomorphism (resp.\ difference ring monomorphism) $\fct{\tau}{\RR}{\seqR}$ is called \notion{$\KK$-homomorphism} (resp.\ \notion{$\KK$-monomorphism} or \notion{$\KK$-embedding}) if for all $c\in\KK$  we have that
$\tau(c)=\langle c,c,\dots\rangle.$

As mentioned already above, our final goal is to construct a $\KK$-monomorphism $\fct{\tau}{\FF[t_1,\dots,t_e]}{\seqR}$. For this task we exploit the following property.

If $\fct{\tau}{\RR}{\seqR}$ is a $\KK$-homomorphism, there is
a map $\fct{\ev}{\RR\times\NN}{\KK}$ with
\begin{equation}\label{Equ:EvDef}
\tau(f)=\langle\ev(f,0),\ev(f,1),\dots\rangle
\end{equation}
for all $f\in\RR$ which has the following properties: For all
$c\in\KK$ there is a $\delta\geq0$ with
\begin{align}\label{Ev:Const}
\forall i\geq\delta:&\;\ev(c,i)=c;\\
\intertext{for all $f,g\in\RR$ there is a $\delta\geq 0$ with}
\forall i\geq\delta:&\;\ev(f\,g,i)=\ev(f,i)\,\ev(g,i),\label{Ev:Mult}\\
\forall i\geq\delta:&\;\ev(f+g,i)=\ev(f,i)+\ev(g,i)\label{Ev:Add};\\
\intertext{and for all $f\in\RR$ and $j\in\ZZ$ there is a $\delta\geq 0$ with}
\label{Ev:Shift}
\forall i\geq\delta:&\;\ev(\sigma^j(f),i)=\ev(f,i+j).
\end{align}
Conversely, if there is a function $\fct{\ev}{\RR\times\NN}{\KK}$
with~\eqref{Ev:Const},~\eqref{Ev:Mult}, \eqref{Ev:Add}
and~\eqref{Ev:Shift}, then the function $\fct{\tau}{\RR}{\seqR}$ defined
by~\eqref{Equ:EvDef} forms a $\KK$-homomorphism.

Subsequently, we assume that a $\KK$-homomorphism/$\KK$-monomorphism is always defined by such a function $\ev$; $\ev$ is also called a \notion{defining function of $\tau$}. To take into account the constructive aspects, we
introduce the following functions for $\ev$.

\begin{definition}
Let $\dfield{\RR}{\sigma}$ be a difference ring and
let $\fct{\tau}{\RR}{\seqR}$ be a $\KK$-homomor\-phism given
by the defining function $\ev$ as in~\eqref{Equ:EvDef}. $\ev$ is called \notion{operation-bounded} by
$\fct{L}{\RR}{\NN}$ if for all $f\in\RR$ and $j\in\ZZ$ with
$\delta=\delta(f,j):=L(f)+\max(0,-j)$ we have~\eqref{Ev:Shift} and
for all $f,g\in\RR$ with $\delta=\delta(f,g):=\max(L(f),L(g))$ we
have~\eqref{Ev:Mult} and~\eqref{Ev:Add}; moreover, we require that for all $f\in\RR$ and all $j\in\ZZ$ we have $L(\sigma^j(f))\leq L(f)+\max(0,-j)$. Such a function is also
called \notion{o-function for $\ev$}. $\ev$ is called \notion{zero-bounded} by
$\fct{Z}{\RR}{\NN}$ if for all $f\in\RR^*$ and all $i\geq Z(f)$ we
have $\ev(f,i)\neq0$; such a function is also called
\notion{z-function for $\ev$}.
\end{definition}

\begin{example}\label{Exp:Embedding1}
Given the \pisiDE-field $\dfield{\KK(x)}{\sigma}$ over $\KK$ with $\sigma(x)=x+1$, we obtain a $\KK$-homomorphism $\fct{\tau}{\KK(x)}{\seqR}$ by taking the defining function~\eqref{Equ:EvalRat}; here we assume that $f=\frac{p}{q}\in\KK(x)$ with $p\in\KK[x]$ and $q\in\KK[x]^*$ are coprime. For the $o$-function $L(f)$ we take the minimal non-negative integer $l$ such that $q(k+l)\neq0$ for all $k\in\NN$, and as $z$-function we take $Z(f)=L(p\,q)$.
Note: Since $p(x)$ and $q(x)$ have only finitely
many roots, $\tau(\frac{p}{q})=\vect{0}$ iff
$\frac{p}{q}=0$. Hence $\tau$ is injective.
\end{example}

Summarizing, the \pisiDE-field $\dfield{\KK(x)}{\sigma}$ with $\sigma(x)=x+1$ can be embedded into $\dfield{\seqR}{\Shift}$. More generally, if $\dfield{\FF}{\sigma}$ is the $q$-mixed difference field,  $\fct{\tau}{\FF}{\seqR}$ with the defining function $\ev$ given in~\eqref{Equ:EvalMixed} is a $\KK$-monomorphism. In addition, there are a computable $o$-function $L$ and a computable $z$-function $Z$ for $\ev$; for details we refer to~\cite[Cor.~4.10]{Schneider:09b} which relies on~\cite{Bauer:99}.

\begin{example}\label{Exp:Embedding2}
Take the rational difference field $\dfield{\KK(x)}{\sigma}$ and the $\KK$-monomor\-phism $\tau$ with defining function $\ev$ and the
$o$-function $L$ from Example~\ref{Exp:Embedding1} and consider the \sigmaDE-extension $\dfield{\KK(x)(h)}{\sigma}$ of $\dfield{\KK(x)}{\sigma}$ with $\sigma(h)=h+\frac{1}{x+1}$.
We get a $\KK$-homomorphism
$\fct{\tau'}{\KK(x)[h]}{\seqP{\KK}}$ where the defining function $\ev'$ is given by
$\ev'(h,k)=H_k$
and\\[-0.2cm]
$$\ev'(\sum_{i=0}^df_ih^i,k)=\sum_{i=0}^d\ev(f_i,k)\ev'(h,k)^i.$$
As $o$-function we can take $L'(\sum_{i=0}^df_ih^i)=\max(L(f_i)|0\leq i\leq d)$. Now suppose that $\tau'$ is not injective. Then we can take
$f=\sum_{i=0}^df_ih^i\in\KK(x)[h]\setminus\{0\}$ with $\deg(f)=d$ minimal
such that $\tau'(f)=\vect{0}$. Since $\tau$ is injective, $f\notin\KK(x)$. Define
$g:=\sigma(f_d)f-f_d\sigma(f)\in\KK(x)[h].$
Note that $\deg(g)<d$ by construction. Moreover,
$$\tau'(g)=\tau(\sigma(f_d))\tau'(f)-
\tau(f_d)\tau'(\sigma(f)).$$
Since $\tau'(f)=\vect{0}$ by assumption and
$\tau'(\sigma(f))=\Shift(\tau'(f))=\Shift(\vect{0})=\vect{0}$, it follows $\tau'(g)=\vect{0}$.
By the minimality of $\deg(f)$, $g=0$, i.e.,
$\sigma(f_d)f-f_d\sigma(f)=0$, or
equivalently,
$\frac{\sigma(f)}{f}=\frac{\sigma(f_d)}{f_d}\in\KK(x).$
As $f\notin\KK(x)$, this contradicts~\cite[Theorem~4]{Karr:81}.
\end{example}

\begin{example}\label{Exp:Embedding3}
Take the rational difference field $\dfield{\KK(x)}{\sigma}$ and the $\KK$-monomor\-phism $\tau$ with defining function $\ev$ and $o$-function $L$ from Example~\ref{Exp:Embedding1}, and consider the \piE-extension $\dfield{\KK(x)(b)}{\sigma}$ of $\dfield{\KK(x)}{\sigma}$ with $\sigma(b)=\frac{x+1}{2(2x+1)}b$.
We get a $\KK$-homomorphism $\fct{\tau'}{\KK(x)[b]}{\seqP{\KK}}$ with its defining function specified by
$\ev'(b,k)=\prod_{i=1}^k\frac{i}{2 (2 i-1)}=\binom{2k}{k}^{-1}$ and~\eqref{Equ:DefineEvExt} where $t:=b$; note that $\tau'(b)$ has no zero entries by construction. We take $L'(\sum_{i=0}^df_ib^i)=\max(L(f_i)|0\leq i\leq d))$ as $o$-function. By similar arguments as in Ex.~\ref{Exp:Embedding2} it follows that $\tau$ is injective.
\end{example}

\noindent More generally, we arrive at the following result; see~\cite{Schneider:09b} for a detailed proof.

\begin{lemma}\label{Lemma:LiftEvToPoly}
Let $\dfield{\FF(t_1)\dots(t_e)(t)}{\sigma}$ be a polynomial
\pisiSE-extension of $\dfield{\FF}{\sigma}$ with
$\KK:=\const{\FF}{\sigma}$ and $\sigma(t)=\alpha\,t+\beta$. Let
$\fct{\tau}{\FF[t_1]\dots[t_e]}{S(\KK)}$ be a $\KK$-monomorphism with a defining function $\ev$ as in~\eqref{Equ:EvDef}; let $L$ be an $o$-function for $\ev$ and let $Z$ be $z$-function for $\ev|_{\FF}$ ($\ev$ is restricted on $\FF$). Then:
\begin{myEnumerate}
\item There is a $\KK$-monomorphism  $\fct{\tau'}{\FF[t_1]\dots[t_e][t]}{S(\KK)}$ with a defining function $\ev'$ such that $\ev'|_{\FF[t_1,\dots,t_e]}=\ev$; if $\beta=0$, $\ev'(t,k)\neq0$ for all $k\geq r$ for some $r\in\NN$. Such a $\tau'$ is uniquely determined by
\begin{equation}\label{Equ:SumProdHom}
\ev'(t,k)=\begin{cases}
\displaystyle c\,\prod_{i=r}^k\ev(\alpha,i-1)&\text{ if $\sigma(t)=\alpha\,t$}\\
\displaystyle\sum_{i=r}^k\ev(\beta,i-1)+c&\text{ if  $\sigma(t)=t+\beta$,}
\end{cases}
\end{equation}
up to the choice of $r\in\NN$ and $c\in\KK$; we require $c\neq0$, if $\beta=0$.
\item Fixing~\eqref{Equ:SumProdHom} we obtain, e.g., the following defining function for $\tau'$:
\begin{equation}\label{Equ:DefineEvExt}
\ev'(\sum_{i=0}^df_it^i,k):=\sum_{i=0}^d\ev(f_i,k)\ev'(t,k)^i\quad\forall k\in\NN.
\end{equation}
\item In particular, there is an $o$-function $L'$ for $\ev'$ with $L'|_{\FF[t_1,\dots,t_e]}=L$; if $L$ and $Z$ are computable, $L'$ can be computed. We can choose (as a constructive example)
\begin{equation}\label{Equ:ChooseLowerBound}
r=\begin{cases}
\max(L(\alpha),Z(\alpha))+1&\text{ if } \sigma(t)=\alpha\,t\\
L(\beta)+1&\text{ if } \sigma(t)=t+\beta.
\end{cases}
\end{equation}
\end{myEnumerate}
\end{lemma}

\begin{remark}\label{Remark:ZImplicit}
Let $\dfield{\FF(t_1)\dots(t_e)}{\sigma}$ be a polynomial
\pisiSE-ext.\ of $\dfield{\FF}{\sigma}$ with $\KK=\const{\FF}{\sigma}$ and let $\fct{\tau}{\FF[t_1]\dots[t_e]}{S(\KK)}$ be a $\KK$-homomorphism with a defining function $\ev$. Then there is implicitly a $z$-function for $\ev|_{\FF}$; see~\cite[Lemma~4.3]{Schneider:09b}.
\end{remark}

Applying Lemma~\ref{Lemma:LiftEvToPoly} iteratively produces the following result.

\begin{theorem}\label{Thm:EmbedTheorem}
Let $\dfield{\FF(y_1)\dots(y_r)(t_1)\dots(t_e)}{\sigma}$ be a polynomial \pisiSE-extension of $\dfield{\FF}{\sigma}$ with
$\KK:=\const{\FF}{\sigma}$; let $\fct{\tau}{\RR}{\seqR}$ with $\RR=\FF[y_1,\dots,y_r]$ be a $\KK$-embedding with a defining function~\eqref{Equ:EvDef} and with an $o$-function $L$.\\
Then there is a
$\KK$-embedding $\fct{\tau'}{\RR[t_1]\dots[t_e]}{\seqR}$ with a defining function $\ev'$ and with an $o$-function $L'$ s.t.\
$\ev'|_{\RR}=\ev$ and $L'|_{\RR}=L$.
\end{theorem}

\noindent This construction that leads in Theorem~\ref{Thm:EmbedTheorem} to $\ev'$ is called \notion{canonical}, if it is performed iteratively as described in~\eqref{Equ:SumProdHom} and~\eqref{Equ:DefineEvExt} of Lemma~\ref{Lemma:LiftEvToPoly}. Note that any defining function $\ev'$ with~\eqref{Equ:SumProdHom} evaluates as~\eqref{Equ:DefineEvExt} if $k$ is chosen big enough; in our canonical construction we assume that~\eqref{Equ:DefineEvExt} holds for all $k\geq0$.

Since $\tau'$ in Lemma~\ref{Lemma:LiftEvToPoly} is uniquely determined by~\eqref{Equ:SumProdHom} the following holds.

\begin{theorem}\label{Thm:EmbedTheorem2}
Let $\dfield{\FF}{\sigma}$ be a \pisiSE-field over $\KK$ and let $\dfield{\FF(y_1)\dots(y_r)}{\sigma}$ be a polynomial \pisiSE-extension of $\dfield{\FF}{\sigma}$. Let $\fct{\tau}{\FF[y_1,\dots,y_r]}{\seqR}$ be a $\KK$-embedding with a defining function $\ev$ together with an $o$-function for $\ev$. Take the measured sequence domain $(\FF,\ev,\delta_{\KK})$.
Then for any $f\in\FF[y_1,\dots,y_r]$, there is an $F\in\ProdSum(\FF)$ s.t.
$\tau(f)=\langle F(k)\rangle_{k\geq0}$ and $\dd(F)=\delta_{\KK}(F)$.
\end{theorem}

\noindent Summarizing, given such a difference ring $\dfield{\FF[y_1,\dots,y_r]}{\sigma}$ and $\KK$-monomorphism, one can rephrase the elements of $\FF[y_1,\dots,y_r]$ as expressions from $\ProdSum(X)$ such that the depth of both domains are identical.

\begin{example}\label{Exp:EmbedH1to4}
Take the \pisiDE-field $\dfield{\QQ(x)}{\sigma}$ with $\sigma(x)=x+1$ together with the $\QQ$-embedding $\fct{\tau}{\QQ(x)}{\seqP{\QQ}}$ with defining function~\eqref{Equ:EvalRat} as carried out in Ex.~\ref{Exp:Embedding1} ($\KK=\QQ$); let $(\Sum(\QQ(x)),\ev,\dd)$ be the sum sequence domain over $\QQ(x)$. Moreover, consider the \pisiDE-field $\dfield{\QQ(x)(h)(h_2)(h_3)(h_4)}{\sigma}$ from Ex.~\ref{Exp:GoodConstruction}. Then we can construct the $\QQ$-embedding $\fct{\tau'}{\QQ(x)[h,h_2,h_3,h_4]}{\seqP{\QQ}}$ with the defining function $\ev'$ which is canonically given by $\ev'|_{\QQ(x)}=\ev$ and by
\begin{equation}\label{Equ:HDFMap}
\begin{aligned}
\ev'(h,k)&=\ev(\Sum(1,\tfrac{1}{x}),k)=H_k,\\
\ev'(h_j,k)&=\ev(\Sum(1,\tfrac{1}{x^j}),k)=H^{(j)}_k\text{ for }j\in\{2,3,4\}.
\end{aligned}
\end{equation}
Note that $\dd(\Sum(1,\tfrac{1}{x}))=\depthQ{h}$ and $\dd(\Sum(1,\tfrac{1}{x^j}))=\depthQ{h_j}$ for $j\in\{2,3,4\}$.
\end{example}

Recall that we want to solve problem~\DOS\ for a measured sequence domain $(X,\ev,\dd)$. In Section~\ref{Sec:OptimalRep} we shall solve this problem for the following general setting.
Loosely speaking, the terms of indefinite nested sums and products $X$ are modeled by polynomials from $\FF[y_1,\dots,y_r]$ and the reinterpretation of the corresponding product-sum expressions is accomplished by its $\KK$-monomorphism from $\FF[y_1,\dots,y_r]$ into the ring of sequences; in particular, the depth of such a product-sum expression is equal to the depth of the corresponding polynomial from $\FF[y_1,\dots,y_r]$.

\begin{example}\label{Exp:Embedding4}
We start as in Example~\ref{Exp:EmbedH1to4}, but now we take the \piE-extension $\dfield{\FF(y_1)\dots(y_r)}{\sigma}$ of $\dfield{\FF}{\sigma}$ such that $\alpha_j=\frac{\sigma(y_j)}{y_j}\in\FF$ for $1\leq j\leq r$. Then we can extend the $\KK$-embedding $\tau$ to $\fct{\tau'}{\FF[y_1,\dots,y_r]}{\seqR}$ with the defining function $\ev'$ canonically given by $\ev'|_{\FF}={\ev}$ and
$$\ev'(y_j,k)=c_j\,\prod_{i=r_j}^k\alpha_j(i)$$
for all $1\leq j\leq r$ with
$r_j\geq Z(\alpha_j)$ and $c_j\in\KK^*$. Note: with $F_j=c_j\Prod(r_j,\alpha_j)$ we have $F_j(k)=\ev'(y_j,k)$ and $\dd(F_j)=\depth{y_j}$. Moreover, we can model a finite set of hypergeometric terms in the sequence domain $(X,\ev',\delta_{\KK})$ with $X:=\FF[y_1,\dots,y_r]$.
\end{example}

Similarly, we are in the position to handle $q$-hypergeometric sequences or mixed hypergeometric sequences. More generally, we can handle the following case.

\medskip

\MyFrame{\begin{gencase}\label{Exp:GeneralSequenceDom}
\textbf{The ground field\footnote{E.g., we can take the $q$-mixed difference field $\dfield{\FF}{\sigma}$ with $\FF=\KK(x,x_1,\dots,x_m)$ from Ex.~\ref{Exp:MixedDF}, and we can take the $\KK$-embedding $\fct{\tau_0}{\FF}{\seqR}$ where $\ev_0:=\ev$ is defined as in~\eqref{Equ:EvalMixed}; note that the measured sequence domain $(\FF,\ev_0,\delta_{\KK})$ has been presented in Ex.~\ref{Exp:MixedTerms}. From the computational point of view we assume that $\KK$ is of the form as stated in Remark~\ref{Remark:KAssum}}.} Let $\dfield{\FF}{\sigma}$ be a \pisiDE-field over $\KK$, let $\fct{\tau_0}{\FF}{\seqR}$ be a $\KK$-embedding with a defining function $\fct{\ev_0}{\FF\times\NN}{\KK}$, and let $\fct{L_0}{\FF}{\NN}$ be an $o$-function and $\fct{Z}{\FF}{\NN}$ be a $z$-function for $\ev_0$; moreover, consider the sequence domain $(\FF,\ev_0,\delta_{\KK})$.

\textbf{A polynomial extension.} In addition, choose a polynomial \pisiDE-extension $\dfield{\FF(y_1)\dots(y_r)}{\sigma}$ of $\dfield{\FF}{\sigma}$ and set $X:=\FF[y_1,\dots,y_r]$. Then extend the $\KK$-embedding $\tau_0$ to $\fct{\tau}{X}{\seqR}$
by extending the defining function $\ev_0$ canonically to $\fct{\ev}{X\times\NN}{\KK}$ and by extending the $o$-function $L_0$ to $L$ following Lemma~\ref{Lemma:LiftEvToPoly}; if $L_0$ is computable, also $L$ is computable. By construction it follows that for $1\leq i\leq r$ there exist $F_i\in\ProdSum(\FF)$ such that
\begin{equation}\label{Equ:ExtSumProdCor}
\ev(y_i,k)=F_i(k)\quad\forall k\geq0\quad\text{and }\dd(F_i)=\depth{y_i}.
\end{equation}
In particular, for each $f\in X$, one gets explicitly an $F\in\ProdSum(X)$ s.t.\ $\ev(f,k)=F(k)$ for all $k\in\NN$ and s.t.\ $\dd(F)=\delta_{\KK}(f)$.\\
\textbf{The sequence domain.} We obtain the sequence domain $(X,\ev,\delta_{\KK})$ which models the product-sum expressions~\eqref{Equ:ExtSumProdCor} with the depth given by $\delta_{\KK}$.
\end{gencase}
}

\section{Combining the steps: Finding optimal nested sum representations}\label{Sec:OptimalRep}

E.g., for the $q$-mixed sequence domain $(\FF,\ev,\dd)$ from Ex.~\ref{Exp:MixedTerms} with $X=\FF=\KK(x,x_1,\dots,x_m)$ we will solve problem~\DOS\ for $A\in\Sum(\FF)$ as follows; here we assume that $\KK$ is of the form as stated in Remark~\ref{Remark:KAssum}

Take the $q$-mixed difference field $\dfield{\FF}{\sigma}$ over $\KK$ with the automorphism $\sigma$ defined in Example~\ref{Exp:MixedDF}. Moreover, take the
the $\KK$-embedding $\fct{\tau}{\FF}{\seqR}$ with the defining function $\ev$ given in~\eqref{Equ:EvalMixed}, and choose a computable $o$-function $L$ and a computable $z$-function $Z$ for $\ev$. Then by Theorem~\ref{Thm:RepresentSequInDepthExt} below the following construction can be carried out algorithmically.

\smallskip

\noindent{\bf Step I: Reduction to a \pisiDE-field.} Given $A\in\Sum(\FF)$,
construct a polynomial \sigmaDE-extension $\dfield{\FF(s_1)\dots(s_u)}{\sigma}$ of $\dfield{\FF}{\sigma}$ and extend the $\KK$-monomorphism $\tau$ to $\fct{\tau'}{\FF[s_1,\dots,s_u]}{\seqR}$ with a defining function $\ev'$ such that the following additional property holds: We can take explicitly an $a\in\FF[s_1,\dots,s_u]$ and a $\lambda\in\NN$ such that\footnotetext{Recall that $A(k)=\ev(A,k)$; here $\ev$ is the evaluation function of the sequence domain $(X,\ev,\dd)$ where $X=\FF$ (or $X=\FF[y_1,\dots,y_r]$ as defined in the General case~\ref{Exp:GeneralSequenceDom}).}
\begin{equation}\label{Equ:AEval}
\ev'(a,k)=A(k)\quad \forall k\geq\lambda\quad\text{and\footnotemark}\;\depth{a}\leq\dd(A).
\end{equation}
Note that we rely on the fact that all our sums are represented in \pisiDE-fields; for general \pisiSE-fields $\depth{a}$ might be bigger than $\dd(A)$, see Example~\ref{Exp:BadConstruction}.\\[0.1cm]
\noindent{\bf Step II: Reinterpretation as a product-sum expression.} In particular, by the concrete construction of the $\KK$-monomorphism based on the iterative application of Lemma~\ref{Lemma:LiftEvToPoly}, construct a $B\in\Sum(\FF)$ such that
\begin{equation}\label{Equ:BEval}
\ev'(a,k)=B(k)\quad\forall k\geq 0\quad\text{and }\delta_{\KK}(a)=\dd(B).
\end{equation}

Then due to the properties of the \pisiDE-field and the fact that $\tau'$ is a $\KK$-mono\-morphism (in particular, that $\tau'$ is injective), we will show in Theorem~\ref{Thm:DepthOptExtIsDepthOpt} that the depth of $B\in\Sum(\FF)$ is $\Sum(\FF)$-optimal, i.e., $B$ together with $\lambda$ are a solution of problem~\DOS.

\medskip

We will solve problem~\DOS\ for the General case~\ref{Exp:GeneralSequenceDom} by applying exactly the same mechanism as sketched above.

\begin{theorem}\label{Thm:RepresentSequInDepthExt}
Let $\dfield{\FF(y_1)\dots(y_r)}{\sigma}$ with $X=\FF[y_1,\dots,y_r]$ be a \pisiDE-field over $\KK$, let $\fct{\tau}{X}{\seqR}$ be $\KK$-embedding  with $\ev$, $L$ and $Z$, and let $(X,\ev,\delta_{\KK})$ be a sequence domain as stated in the General case~\ref{Exp:GeneralSequenceDom}; in particular let $(\Sum(X),\ev,\dd)$ be the sum sequence domain over $X$. Then for any $A\in\Sum(X)$
there is a \sigmaDE-extension $D:=\dfield{\FF(y_1)\dots(y_r)(s_1)\dots(s_u)}{\sigma}$
of $\dfield{\FF(y_1)\dots(y_r)}{\sigma}$, where $D$ is a polynomial extension of $\dfield{\FF}{\sigma}$, and there is a $\KK$-embedding
\begin{equation}\label{Equ:KEmbed}
\fct{\tau'}{X[s_1,\dots,s_u]}{\seqR}
\end{equation}
where the defining function $\ev'$ and its $o$-function $L'$ are extended from $X$ to $X[s_1,\dots,s_u]$, with the following property: There are $\lambda\in\NN$ and $a\in\RR$ such that~\eqref{Equ:AEval}; in particular, for any $h\in X[s_1,\dots,s_u]$ there is an $H\in\Sum(X)$ s.t.
\begin{equation}\label{Equ:EmbedInConstruction}
\ev'(h,k)=H(k)\quad\forall k\geq0\quad\text{and }\depth{h}=\dd(H).
\end{equation}
This extension, the defining function $\ev'$ for $\tau'$, $\lambda$, and $a$ can be given explicitly, if $L$ and $Z$ are computable and if $\KK$ has the form as stated in Remark~\ref{Remark:KAssum}.
\end{theorem}
\begin{proof}
We show the theorem by induction on the depth. If $A\in\Sum(X)$ with $\dd(A)=0$, then $A\in\KK$ and the statement clearly holds. Now suppose that we have shown the statement for expressions with depth$\leq d$ and take $A\in\Sum(X)$ with $\dd(A)=d+1$. Let $A_1,\dots,A_l$ be exactly those subexpressions of $A$ which do not occur inside of a sum  and which cannot be split further by $\oplus$ and $\otimes$, i.e., for $1\leq i\leq l$, either $A_i\in X$ or $A_i$ is a sum. First we consider $A_1$.
If $A_1\in X$, then for $r_1=0$, $a_1=A_1$, $\ev'=\ev$ and $i=1$ we have
\begin{equation}\label{Equ:A1a}
\ev'(a_i,k)=A_i(k)\quad\forall k\geq r_i,\quad\text{and }\depth{a_i}\leq\dd(A_i).
\end{equation}
Moreover, the property~\eqref{Equ:EmbedInConstruction} for any $h\in X$ holds by choosing $H:=h$.\\
Otherwise, $A_1=\Sum(\lambda_1,F_1)$ for some $\lambda_1\in\NN$ and $F_1\in\Sum(X)$ with $\dd(F_1)\leq d$.
If $\dd(A_1)\leq d$, we get by induction a \sigmaDE-extension $D:=\dfield{\FF(y_1)\dots(y_r)(s_1)\dots(s_u)}{\sigma}$ of $\dfield{\FF(y_1)\dots(y_r)}{\sigma}$ where $D$ is a polynomial extension of $\dfield{\FF}{\sigma}$; moreover we can extend $\tau$ to a $\KK$-embedding $\fct{\tau'}{\DD}{\seqR}$ with $\DD:=X[s_1,\dots,s_u]$ where its defining function $\ev'$ is extended from $X$ to $\DD$, and we can extend the $o$-function $L$ to an $o$-function $L'$ for $\ev'$ s.t.\ the following holds: There are $a_1\in\DD$ and $r_1\in\NN$ s.t.\ for $i=1$ we have~\eqref{Equ:A1a}; in particular, for any $h\in\DD$, there is an $H\in\Sum(X)$ s.t.~\eqref{Equ:EmbedInConstruction}.\\
If $\dd(A_1)=d+1$, we can take by the same reasoning such an extension $D$ of $\dfield{\FF}{\sigma}$ with $\DD:=X[s_1,\dots,s_u]$ and $\tau'$ with a defining function $\ev'$ together with an $o$-function $L'$ in which we can take $f_1\in\DD$ and $l_1\in\NN$ s.t.
$$\ev'(f_1,k)=F_1(k)\quad\forall k\geq l_1,\quad\text{and }
\depth{f_1}\leq\dd(F_1).$$
By Theorem~\ref{Thm:ConstructDeltaSigma} take a \sigmaDE-extension $E:=\dfield{\FF(y_1)\dots(y_r)(s_1)\dots(s_u)(t_1)\dots(t_v)}{\sigma}$ of $\dfield{\FF(y_1)\dots(y_r)(s_1)\dots(s_u)}{\sigma}$ such that $E$ a polynomial extension of $\dfield{\FF}{\sigma}$ and in which we have $g\in\EE$ with $\EE:=\DD[t_1,\dots,t_v]$ such that
$$\sigma(g)=g+\sigma(f_1).$$
Moreover, by iterative application of Lemma~\ref{Lemma:LiftEvToPoly} we can extend the $\KK$-embedding $\tau'$ from $\DD$ to a $\KK$-embedding $\fct{\tau'}{\EE}{\seqR}$ by extending the defining function $\ev'$ canonically from $\DD$ to $\EE$, and we can extend $L'$ to an $o$-function for $\ev'$;
note that this construction can be performed such that for any $h\in\EE$ there is $H\in\Sum(X)$ with~\eqref{Equ:EmbedInConstruction}. Now take\footnote{Remark: If $f_1$ is a hypergeometric term, $r_1$ can be also obtained by analyzing only $f_1$ (without knowing the telescoping solution $g$); for more details see~\cite{Abramov:05b}.} $r_1:=\max(l_1,L'(f_1),L'(g)+1)$ and define $c:=\sum_{k=l_1}^{r_1-1} F(k)-\ev'(g,r_1-1)\in\KK$. Then for all $n\geq r_1$,
\begin{multline*}
\ev'(g+c,n)=\ev(\sigma^{-1}(g)+f_1,n)+c=\ev'(g,n-1)+\ev'(f_1,n)+c\\
=\ev'(g,n-1)+F_1(n)+c=\dots=\ev'(g,r_1-1)+\sum_{k=r_1}^{n}F_1(k)+c=A_1(n).
\end{multline*}
Set $a_1:=g+c\in\EE$. Then~\eqref{Equ:A1a} for $i=1$. Moreover,
since $\depth{g}\leq\depth{\sigma(f_1)}+1$ by Thm.~\ref{Thm:DepthStable}, $$\depth{a_1}=\depth{g}\leq\depth{\sigma(f_1)}+1=\depth{f_1}+1\leq \dd(F_1)+1\leq d+1.$$
We continue to consider $A_2,\dots,A_l$ and finally arrive at a polynomial \sigmaDE-extension $\dfield{\FF(y_1)\dots(y_r)(s_1)\dots(s_u)(t_1)\dots(t_{v'})}{\sigma}$ of $\dfield{\FF}{\sigma}$; during this construction we can extend $\tau$ to a $\KK$-embedding $\fct{\tau'}{\RR}{\seqR}$ with $\RR:=X[s_1,\dots,s_u][t_1,\dots,t_{v'}]$ and with a defining function $\ev'$ and we can extend $L$ to an $o$-function $L'$ for $\ev'$ s.t. the following holds. We can take $a_1,\dots,a_l\in\RR$ and $r_1,\dots,r_l\in\NN$ s.t.\ for $1\leq i\leq l$ we have~\eqref{Equ:A1a}; in particular, for any $h\in\RR$, there is an $H\in\Sum(X)$ s.t.~\eqref{Equ:EmbedInConstruction}.\\
Finally, we construct $a\in\RR$ by applying in the given $A$ the substitution $A_i\to a_i$ for all $1\leq i \leq l$ and by replacing $\otimes$ and $\oplus$ with the field operations $\cdot$ and $+$, respectively. Then it follows that~\eqref{Equ:AEval} for $\lambda:=\max(r_1,\dots,r_l,L'(a_1),\dots,L'(a_l))$. This completes the induction step. Note that all the construction steps can be carried out by algorithms if $L$ and $Z$ are computable and if $\KK$ has the form as stated in Remark~\ref{Remark:KAssum}. In particular, $\ev'$, $a$ and $\lambda$ can be given explicitly.
\end{proof}

\begin{example}\label{Exp:GoodConstrLinkObj}
For the input sum~\eqref{Equ:NestedSumExp} the presented procedure in Theorem~\ref{Thm:RepresentSequInDepthExt} carries out simultaneously the constructions from Examples~\ref{Exp:GoodConstruction} and~\ref{Exp:EmbedH1to4}: We obtain the \pisiDE-field $\dfield{\QQ(x)(h)(h_2)(h_3)(h_4)}{\sigma}$ over $\QQ$ together with the $\QQ$-embedding $\fct{\tau'}{\QQ(x)[h,h_2,h_3,h_4]}{\seqP{\QQ}}$ where the defining function $\ev'$ is canonically given by $\ev'|_{\QQ(x)}=\ev$ for $\ev$ as in~\eqref{Equ:EvalRat} and by~\eqref{Equ:HDFMap}. By construction, we can link the sums in~\eqref{Equ:SumOrder} with $h,s',h_2,t',a'\in\QQ(x)[h,h_2,h_3,h_4]$ from Example~\ref{Exp:GoodConstruction}: for $n\in\NN$,
\begin{equation}\label{SeqDepthOpt2}
\begin{aligned}
\ev'(h,n)&=H_n,&\ev'(s',n)&=S(n),\\
\ev'(h_2,n)&=H^{(2)}_n,
&\ev'(t',n)&=T(n),&\ev'(a',n)&=A(n).
\end{aligned}
\end{equation}
The depths of $H_k,S,H^{(2)}_k,T,A$ are $2,3,2,3,4$, respectively. The corresponding depths  $\depthQ{h}=\depthQ{s'}=\depthQ{h_2}=\depthQ{t'}=\depthQ{a'}=2$ in the \pisiDE-field are the same or have been improved. Using~\eqref{Equ:HDFMap} we can reinterpret, e.g, $s'$ as the sum expression $F:=\frac{1}{2}(\Sum(1,1/x)^2+\Sum(1,1/x^2))\in\Sum(\QQ(x))$ with $F(n)=S(n)$ for all $n\geq0$ and $\dd(F)=\depthQ{s'}$; this leads to the identity $s(n)=\frac{1}{2}(H_n^2+H^{(2)}_n)$.
In the same way we obtain sum expressions in $\Sum(\QQ(x))$ for the \pisiDE-field elements
$h,h_2,t',a'$ in~\eqref{SeqDepthOpt2}, and we arrive at the following identities. For $n\geq0$,
\begin{equation*}
\begin{aligned}
H_n&=H_n,& s(n)&=\tfrac{1}{2}(H_n^2+H^{(2)}_n),\\
H^{(2)}_n&=H^{(2)}_n,&
T(n)&=\tfrac{1}{3} \left(H_n^3+3 H_n^{(2)} H_n+2H_n^{(3)}\right),& A(n)&=B(n)
\end{aligned}
\end{equation*}
\noindent where $B$ is given as in~\eqref{Equ:BSol}.
\end{example}

We remark that this translation mechanism presented in Theorem~\ref{Thm:RepresentSequInDepthExt} is implemented in the summation package~\SigmaP. Namely, given the General case~\ref{Exp:GeneralSequenceDom} ($\KK$ as stated in Remark~\ref{Remark:KAssum}) and given\footnote{In~\SigmaP\ $A$ is inserted, e.g., in the form~\eqref{Equ:NestedSumExp} without using evaluation functions like \eqref{Equ:EvalRat} or~\eqref{Equ:EvalMixed}; this implies that the lower bounds of the involved sums and products must be chosen in such a way that no zeros occur in the denominators during any evaluation.} $A\in\Sum(X)$, \SigmaP\ computes the following ingredients:

\begin{myItemize}
\item A \sigmaDE-extension $\dfield{\FF(y_1)\dots(y_r)(s_1)\dots(s_u)}{\sigma}$
of $\dfield{\FF(y_1)\dots(y_r)}{\sigma}$ and a $\KK$-em\-bedding~\eqref{Equ:KEmbed} with a defining function $\ev'$ and $o$-function $L'$.
\item $a\in X[s_1,\dots,s_u]$ and $\lambda\in\NN$ such that~\eqref{Equ:AEval}.
\item $\tau'(a)$ is given explicitly by a\footnote{In \SigmaP\ the lower bounds of the sums and products are computed by~\eqref{Equ:ChooseLowerBound}. Looking closer at this construction, no zeros occur in the involved denominators of $B$ when performing the evaluation $B(n)$ for $n\geq\lambda$. Hence the output can be returned, e.g., in the form like~\eqref{Equ:NestedSumExp} or~\eqref{Equ:BSol} which is free of any explicit evaluation functions like~\eqref{Equ:EvalRat} or~\eqref{Equ:EvalMixed}.} $B\in\Sum(X)$ s.t.~\eqref{Equ:BEval}.
\end{myItemize}

\noindent Then~\SigmaP\ outputs for a given $A\in\Sum(X)$ the result $B\in\Sum(X)$ with $\lambda$.

\medskip

The final goal is to prove Theorem~\ref{Thm:DepthOptExtIsDepthOpt} which guarantees that the output $B$ is indeed a solution of problem~\DOS. We start with the following

\begin{lemma}\label{Lemma:EmbedConstr}
Let $\dfield{\FF(y_1)\dots(y_r)}{\sigma}$ be a \pisiDE-field over $\KK$ and let $\tau$ be a $\KK$-mono\-morphism with $\ev$ and $L$ as in the General case~\ref{Exp:GeneralSequenceDom}; let $\dfield{\FF(t_1)\dots(t_e)}{\sigma}$ be a polynomial \sigmaSE-extension of $\dfield{\FF}{\sigma}$ and let $\fct{\rho}{\FF[t_1,\dots,t_e]}{\seqR}$ be a $\KK$-embedding with a defining function $\ev_{\rho}$ and an $o$-function $L_{\rho}$ such that $\ev_{\rho}|_{\FF}=\ev$.\\
Then
there is a \sigmaSE-extension $D:=\dfield{\FF(y_1)\dots(y_r)(z_1)\dots(z_l)}{\sigma}$ of $\dfield{\FF(y_1)\dots(y_r)}{\sigma}$ where $D$ is a polynomial \sigmaSE-extension of $\dfield{\FF}{\sigma}$ with the following properties:
\begin{myEnumerate}
\item There is a difference field monomor.\ $\fct{\phi}{\FF(t_1,\dots,t_e)}{\FF(y_1,\dots,y_r)(z_1,\dots,z_l)}$
    such that for all $a\in\FF(t_1,\dots,t_e)$,
\begin{equation}\label{Equ:DepthOpt}
\depth{\phi(a)}\leq\depth{a}
\end{equation}
and such that for all $a\in\FF[t_1,\dots,t_e]$,
\begin{equation}\label{Equ:DepthOptPoly}
\phi(a)\in\FF[y_1,\dots,y_r][z_1,\dots,z_l].
\end{equation}

\item There is a $\KK$-embed.\ $\fct{\tau'}{\FF[y_1,\dots,y_r][z_1,\dots,z_l]}{\seqR}$ with a defining function $\ev'$ and an $o$-function $L'$ s.t.\ $\ev'|_{\FF[y_1,\dots,y_r]}=\ev$ and s.t.\ for all $a\in\FF[t_1,\dots,t_e]$,
\begin{equation}\label{Equ:EvEqual}
\tau'(\phi(a))=\rho(a).
\end{equation}
\end{myEnumerate}
\end{lemma}
\begin{proof}
The base case $e=0$ holds with $\phi(a)=a$ for all $a\in\FF$ and $\tau':=\rho$. Suppose the lemma holds for $e$ extensions $\dfield{\HH}{\sigma}$ with $\HH=\FF(t_1)\dots(t_e)$ and let $\dfield{\DD}{\sigma}$ with $\DD:=\FF(y_1)\dots(y_r)(z_1)\dots(z_l)$, $\tau'$ with $\ev'$ and $L'$, $\rho$ with $\ev_{\rho}$, and $\phi$  as stated above; set $\EE=\FF[y_1,\dots,y_r,z_1,\dots,z_l]$. Now let $\dfield{\HH(t)}{\sigma}$ be a \sigmaSE-ext.\ of $\dfield{\HH}{\sigma}$ with $f:=\sigma(t)-t\in\FF[t_1\dots,t_e]$, and take a $\KK$-embedding $\fct{\rho'}{\FF[t_1,\dots,t_e][t]}{\seqR}$ with a defining function $\rho'$ and an $o$-function $L_{\rho'}$ s.t.\ $\ev_{\rho'}|_{\FF[t_1,\dots,t_e]}=\ev_{\rho}$.\\ 
\textbf{Case 1:} If there is no $g\in\DD$ such that
\begin{equation}\label{Equ:LambaTele}
\sigma(g)-g=\phi(f),
\end{equation}
we can take the \sigmaSE-extension $\dfield{\DD(y)}{\sigma}$ of $\dfield{\DD}{\sigma}$ with $\sigma(y)=y+\phi(f)$ by Theorem~\ref{Thm:PiSigma}.1 and we can define a difference field
monomorphism $\fct{\phi'}{\HH(t)}{\EE(y)}$ s.t.\
$\phi'(a)=\phi(a)$ for all $a\in\HH$ and such that $\phi'(t)=y$.
By construction, $\depth{y}=\depth{\phi(f)}+1$. Since
\begin{equation}\label{Equ:SummandIneq}
\depth{\phi(f)}+1\leq\depth{f}+1=\depth{t},
\end{equation}
$\depth{\phi(a)}\leq\depth{a}$ for all $a\in\HH(t)$.
Moreover, since $\phi(f)\in\EE$, it follows that $\dfield{\EE(y)}{\sigma}$ is a polynomial extension of $\dfield{\FF}{\sigma}$. Moreover, for all $a\in\FF[t_1,\dots,t_e,t]$, $\phi(a)\in\EE[y]$. This proves part~(1).\\
Now we extend the $\KK$-embedding $\tau'$ from $\EE$ to $\fct{\tau'}{\EE[y]}{\seqR}$ with the defining function $\ev'$ where $\ev'(f,k)=\ev(f,k)$ for all $f\in\EE$ and $\ev'(y,k)$ is defined as in the right hand side of~\eqref{Equ:SumProdHom}; here we choose
$\beta=\phi(f)$ and $c=\ev_{\rho'}(t,r-1)$ for some $r\in\NN$ properly chosen. In particular, we can extend the $o$-function $L'$ for our extended $\ev'$ by Lemma~\ref{Lemma:LiftEvToPoly} (note that there is a $z$ function for $ev'$ restricted on $\FF$ by Remark~\ref{Remark:ZImplicit}). Then for all $k\geq r$ ($r$ is chosen big enough with $L'_{\rho}$ and $L'$),
\begin{equation}\label{Equ:RNotSpecified}
\begin{split}
\ev'(\phi'(t),k)&=\ev'(y,k)=\sum_{i=r}^k\ev'(\phi(f),i-1)+\ev_{\rho'}(t,r-1)\\
&=\sum_{i=r}^k\ev_{\rho}(f,i-1)+\ev_{\rho'}(t,r-1)=\sum_{i=r+1}^k\ev_{\rho}(f,i-1)+h(r)
\end{split}
\end{equation}
with $h(r)=\ev_{\rho}(f,r-1)+\ev_{\rho'}(t,r-1)=\ev_{\rho'}(f+t,r-1)=\ev_{\rho'}(\sigma(t),r-1)
=\ev_{\rho'}(t,r)$. Applying this reduction $k-r+1$ times shows that
\begin{align*}
\ev'(\phi'(t),k)&=\sum_{i=r+1}^k\ev_{\rho}(f,i-1)+\ev_{\rho'}(t,r)=\dots=\ev_{\rho'}(t,k).
\end{align*}
Hence $\tau'(\phi'(t))=\rho'(t)$, and thus $\tau'(\phi'(a))=\rho'(a)$ for all $a\in\FF[t_1\dots,t_e][t]$.\\
\textbf{Case 2:} Otherwise, if there is a $g\in\DD$ s.t.~\eqref{Equ:LambaTele}, then $g\in\EE$ by Theorem~\ref{Thm:PolyClosure}. In particular, $\depth{g}\leq\depth{\phi(f)}+1$
by Theorem~\ref{Thm:DepthStable}. With~\eqref{Equ:SummandIneq}, it follows that
\begin{equation}\label{Equ:TgIneq}
\depth{g}\leq\depth{t}.
\end{equation}
Now observe that
$$\Shift(\rho'(t))-\rho'(t)=\rho'(\sigma(t)-t)=\rho(f)=\tau'(\phi(f))=\tau'(\sigma(g)-g)=\Shift(\tau'(g))-\tau'(g).$$
Hence $\Shift(\rho'(t)-\tau'(g))=\rho'(t)-\tau'(g)$, i.e., $\rho'(t)-\tau'(g)$ is a constant in $\seqR$. Thus, $\rho'(t)=\tau'(g)+\langle c\rangle_{n\geq0}$ for some $c\in\KK$. Since
$\dfield{\phi(\HH)(g)}{\sigma}$ is a difference field (it is a
sub-difference field of $\dfield{\DD}{\sigma}$), $g$ is
transcendental over $\phi(\HH)$ by Theorem~\ref{Thm:PiSigma}.1. In
particular, we can define the difference field monomorphism
$\fct{\phi'}{\HH(t)}{\DD}$ with $\phi'(a)=\phi(a)$ for all
$a\in\HH$ and $\phi'(t)=g+c$.
Since $g\in\EE$, $\phi'(t)\in\EE$, and therefore $\phi(a)\in\EE$ for all $a\in\FF[t_1,\dots,t_e][t]$. With~\eqref{Equ:TgIneq} and our
induction assumption it follows that $\depth{\tau'(a)}\leq\depth{a}$
for all $a\in\HH(t)$. This proofs part~(1). Note that
$\tau'(\phi'(t))=\tau'(g)+c=\rho'(t)$ by construction. Hence, $\tau'(\phi'(a))=\rho(a)$ for all $a\in\FF[t_1\dots,t_e][t]$. This proves part~(2) and completes the induction step.
\end{proof}

\noindent Note that the proof of Lemma~\ref{Lemma:EmbedConstr} is constructive, if the underlying $o$- and $z$-functions of $\tau$ are computable and if $\KK$ is given as in Remark~\ref{Remark:KAssum}. For simplicity, ingredients like $r_1$ (needed, e.g., for~\eqref{Equ:RNotSpecified}) have not been specified explicitly.

\smallskip

In Theorem~\ref{Thm:SequDepthOpt} we can show the following:
The nested depth of an element in a \pisiDE-field over $\KK$ is smaller or equal to the depth of an element in a \pisiSE-field over $\KK$ provided that both elements can be mapped to the same sequence $\vect{s}\in\seqR$ by appropriate $\KK$-monomorphisms.

\begin{theorem}\label{Thm:SequDepthOpt}
Let $\dfield{\FF(y_1)\dots(y_r)}{\sigma}$ be a \pisiDE-field over $\KK$ and let $\tau$ be a $\KK$-embedding with $\ev$ and $L$ as in the General case~\ref{Exp:GeneralSequenceDom}.  Let $\dfield{\FF(t_1)\dots(t_e)}{\sigma}$ be a polynomial \sigmaSE-extension of $\dfield{\FF}{\sigma}$ and let $\fct{\rho}{\FF[t_1,\dots,t_e]}{\seqR}$ be a $\KK$-embed\-ding with a defining function $\ev_{\rho}$ and an $o$-function for $\ev_{\rho}$. Then for any $\vect{s}\in\tau(\FF[y_1,\dots,y_r])\cap\rho(\FF[t_1,\dots,t_e])$ we have
$$\depth{\tau^{-1}(\vect{s})}\leq\depth{\rho^{-1}(\vect{s})}.$$
\end{theorem}

\begin{proof}
Take a \sigmaSE-extension
$\dfield{\FF(y_1)\dots(y_r)(z_1)\dots(z_l)}{\sigma}$ of $\dfield{\FF(y_1)\dots(y_r)}{\sigma}$, a mo\-nomorphism $\fct{\phi}{\FF(t_1)\dots(t_e)}{\FF(y_1)\dots(y_r)(z_1)\dots(z_l)}$
s.t.\ \eqref{Equ:DepthOpt} and \eqref{Equ:DepthOptPoly} for all $a\in\FF[t_1,\dots,t_e]$, and take a $\KK$-embedding $\fct{\tau'}{\FF[y_1,\dots,y_r,z_1,\dots,z_l]}{\seqR}$ with $\tau'|_{\FF[y_1,\dots,y_r]}=\tau$ s.t.~\eqref{Equ:EvEqual} for all $a\in\FF[t_1,\dots,t_e]$; this is possible by Lemma~\ref{Lemma:EmbedConstr}. Now let $\vect{s}\in\tau(\FF[y_1,\dots,y_r])\cap\rho(\FF[t_1,\dots,t_e])$, and set $f:=(\tau')^{-1}(\vect{s})=\tau^{-1}(\vect{s})\in\FF[y_1,\dots,y_r]$ and $g:=\rho^{-1}(\vect{s})\in\FF[t_1\dots,t_e]$. To this end, define $g':=\phi(g)\in\FF[y_1,\dots,y_r,z_1,\dots,z_l]$. Then by~\eqref{Equ:DepthOpt} we have $\depth{g'}\leq\depth{g}$. Since $\tau'(g')=\rho(g)=\vect{s}$ and $\tau'(f)=\vect{s}$, and since $\tau'$ is injective, $g'=f$. Thus, $\depth{f}=\depth{g'}\leq\depth{g}$.
\end{proof}

Finally, Theorem~\ref{Thm:DepthOptExtIsDepthOpt} shows that the constructed $B\in\Sum(X)$ with~\eqref{Equ:BEval} is indeed a solution of problem~\DOS.

\begin{theorem}\label{Thm:DepthOptExtIsDepthOpt}
Let $\dfield{\FF(y_1)\dots(y_r)}{\sigma}$ with $X=\FF[y_1,\dots,y_r]$ be a \pisiDE-field over $\KK$, let $\fct{\tau}{X}{\seqR}$ be $\KK$-embedding  with $\ev$ and $L$, and let $(X,\ev,\delta_{\KK})$ be a sequence domain as stated in the General case~\ref{Exp:GeneralSequenceDom};
let $D:=\dfield{\FF(y_1)\dots(y_r)(s_1)\dots(s_u)}{\sigma}$ be a \sigmaDE-extension
of $\dfield{\FF(y_1)\dots(y_r)}{\sigma}$ where $D$ is a polynomial extension of $\dfield{\FF}{\sigma}$ and let~\eqref{Equ:KEmbed} be a $\KK$-embedding with a defining function and $o$-function. Moreover, let $A\in\Sum(X)$ and $a\in\RR$ such that $\tau(a)=\langle A(k)\rangle_{k\geq0}$. Then $\depth{a}$ is the $\Sum(X)$-optimal depth of~$A$.
\end{theorem}

\begin{proof}
Take an expression of $\Sum(X)$ that produces $\vect{s}:=\langle A(k)\rangle_{k\geq0}$ from a certain point on and that has minimal depth, say $d$. By Theorem~\ref{Thm:RepresentSequInDepthExt} we can take a \sigmaDE-extension $D:=\dfield{\FF(y_1)\dots(y_r)(t_1)\dots(t_e)}{\sigma}$ of $\dfield{\FF(y_1)\dots(y_r)}{\sigma}$ s.t.\ $D$ is a polynomial extension of $\dfield{\FF}{\sigma}$ and we can assume that there is a $\KK$-embedding $\fct{\rho}{X[t_1,\dots,t_e]}{\seqR}$ with a defining function and an $o$-function with the following property. There is an $a'\in X[t_1,\dots,t_e]$ s.t.\ $\rho(a')=\vect{s}$ and $\depth{a'}\leq d$. By Theorem~\ref{Thm:SequDepthOpt} (applied twice), $\depth{a}=\depth{a'}$. Moreover, $\rho(a')$ and $\tau(a)$ can be defined by elements from $\Sum(X)$ with depth $\depth{a}$ by Theorem~\ref{Thm:EmbedTheorem2}. Since $d$ is minimal, $\depth{a}=d$.
\end{proof}

\section{Application: Simplification of d'Alembertian solutions}\label{Sec:dAlembert}

The d'Alembertian solutions~\cite{Noerlund,Abramov:94,Schneider:01a}, a subclass of Liouvillian
solutions \cite{Singer:99}, of a given recurrence relation are computed by factorizing the recurrence into linear right hand factors as much as possible. Given this factorization, one can read of the d'Alembertian solutions which are of the form
\begin{equation}\label{DefOfDAlembertSol}
h(n)\sum_{k_1=c_1}^nb_1(k_1)\sum_{k_2=c_2}^{k_2}
b_2(k_2)\dots\sum_{k_s=c_s}^{k_{s-1}}b_s(k_s)
\end{equation}
for lower bounds $c_1,\dots,c_s\in\NN$;
here the $b_i(k_i)$ and $h(n)$ are given by the objects  form the coefficients of the recurrence or by products over such
elements. Note that such solutions can be represented in \pisiDE-fields if the occurring products can be rephrased accordingly in \piE-extensions. Then applying our refined algorithms to such solutions~\eqref{DefOfDAlembertSol}, we can find sum representations with minimal nested depth. Typical examples can be found, e.g., in~\cite{Schneider:06c,Schneider:07a,Schneider:07i,Schneider:08e,Schneider:09a,Schneider:T09a,Schneider:09d}.

In the following we present two examples with detailed computation steps that have been provided by the summation package~\SigmaP.

\subsection{An example from particle physics}

In massive higher
order calculations of Feynman diagrams~\cite{Schneider:07h} the following task of simplification arose. Find an alternative sum expression of the definite sum
\begin{equation}\label{Equ:FeynmanSum}
S(n)=\sum_{i=1}^{\infty}\frac{H_{i+n}^2}{i^2}
\end{equation}
such that the parameter $n$ does not occur inside of any summation quantifier and such that the arising sums are as simple as possible. In order to accomplish this task, \SigmaP\ computes in a first step the recurrence relation
\begin{multline*}
-(n+2)(n+1)^3 \left(n^2+7 n+16\right)
   S(n)\\
+(n+2) \left(5n^5+62 n^4+318 n^3+814 n^2+1045
   n+540\right) S(n+1)\\
-2 \left(5
   n^6+84 n^5+603 n^4+2354 n^3+5270
   n^2+6430 n+3350\right)S(n+2)\\
+2 \left(5 n^6+96
   n^5+783 n^4+3478 n^3+8906 n^2+12530
   n+7610\right) S(n+3)\\
-(n+4)\left(5 n^5+88 n^4+630 n^3+2318
   n^2+4453 n+3642\right)
   S(n+4)\\
+(n+4) (n+5)^3\left(n^2+5 n+10\right)
   S(n+5)=-\frac{4 (n+7)}{(n+3) (n+4)}H_n\\
-\frac{2 \left(2 n^7+35 n^6+235 n^5+718
   n^4+824 n^3-283 n^2-869
   n+10\right)}{(n+1) (n+2) (n+3)^2
   (n+4)^2 (n+5)}
\end{multline*}
by a generalized version~\cite{Schneider:07a} of Zeilberger's creative telescoping~\cite{Zeilberger:91}.
Given this recurrence, \SigmaP\ computes the d'Alembertian solutions
\begin{align*}
A_1&=1,\quad A_2=H_n,\quad A_3=H_n^2,\\ A_4&=\sum_{i=2}^n\frac{\displaystyle\sum_{j=2}^i\frac{\displaystyle(2 j-1)
   \sum_{k=1}^j\frac{1}{(2 k-3) (2
   k-1)}}{(j-1)j}}{i},\\
A_5&=\sum_{i=3}^n\frac{\displaystyle\sum_{j=3}^i\frac{\displaystyle(2 j-1)
   \sum_{k=3}^j\frac{\displaystyle2 (k-2) (k-1)
   k H_k-(2 k-1) \left(3 k^2-6
   k+2\right)}{(k-2) (k-1) k (2 k-3) (2
   k-1)}}{(j-1)
   j}}{i},\\
B&=\sum_{i=4}^n\frac{\displaystyle\sum_{j=4}^i\frac{\displaystyle(2 j-1)
   \sum_{k=4}^j\frac{\displaystyle\sum_{l=4}^k\frac{\displaystyle(2 l-3) \left(l^2-3
   l+6\right) \tilde{B}(l)}{(l-3)
   (l-2) (l-1) l}}{(2
   k-3) (2 k-1)}}{(j-1)j}}{i}
\end{align*}
where
$$\tilde{B}=\sum_{r=3}^l-\tfrac{2
   \left(2 r^6-27 r^5+117 r^4-254 r^3+398
   r^2+2 (r-3) (r-2) (r-1) (r+2) H_r
   r-446 r+204\right)}{(r-2) (r-1) r
   \left(r^2-5 r+10\right) \left(r^2-3
   r+6\right)}.$$
To be more precise, $A_i\in\Sum_n(\QQ(x))$, $1\leq i\leq 5$, are the five linearly independent solutions of the homogeneous version of the recurrence, and $B\in\Sum_n(\QQ(x))$ is  one particular solution of the recurrence itself; the depths of $A_1,\dots,A_5,B$ are $0,2,2,4,5,7$, respectively. As a consequence, we obtain the general solution
\begin{equation}\label{Equ:LinComb}
G:=B+c_1 A_1+c_2 A_2+c_3 A_3+c_4 A_4+c_5 A_5
\end{equation}
for constants $c_i$. Checking initial values shows that we have to choose\footnote{$\zeta_k$ denotes the Riemann zeta function at $k$; e.g., $\zeta_2=\pi^2/6$.}
$$c_1=\tfrac{17}{10}\zeta_2^2,\,c_2=\tfrac{1}{12} (48
   \zeta_3-67),\,c_3=\tfrac{31}{12},\,c_4=\tfrac{1}{4} (23-8\zeta_2),\,c_5=-\tfrac{1}{2}$$
in order to match~\eqref{Equ:LinComb} with $S(n)=G(n)$ for all $n\in\NN$.

Finally, \SigmaP\ simplifies the derived expressions further and finds sum representations with minimal nested depth (see problem~\DOS). Following the approach described in the previous sections, it computes the \pisiDE-field $\dfield{\QQ(x)(h)(h_2)(h_4)(H)}{\sigma}$ with $\sigma(x)=x+1$ and
\begin{equation*}
\sigma(h)=h+\tfrac{1}{x+1},\; \sigma(h_2)=h_2+\tfrac{1}{(x+1)^2},\;
\sigma(h_4)=h_4+\tfrac{1}{(x+1)^4},\; \sigma(H)=H+\tfrac{\sigma(h)}{(x+1)^2};
\end{equation*}
in addition, it delivers a $\QQ$-embedding $\fct{\tau}{\QQ(x)[h,h_2,h_4,H]}{\seqQ}$ with the defining function $\fct{\ev}{\QQ(x)[h,h_2,h_4,H]\times\NN}{\QQ}$ canonically given by~\eqref{Equ:EvalRat} for all $f\in\QQ(x)$ and by
\begin{align*}
\ev(h,n)=H_n,&& \ev(h_2,n)=H^{(2)}_n,&& \ev(h_4,n)=H^{(4)}_n,&& \ev(H,n)=\sum_{k=1}^n\frac{H_k}{k^2}.
\end{align*}
Moreover, it finds
\begin{align*}
a_1&=1,\quad a_2=h,\quad
a_3=h^2,\quad a_4=\frac{1}{2}\left(h_2-h^2\right),\quad
a_5=\frac{1}{2} \left(-h^2+2h_2\,h-h\right),\\
b&=\frac{1}{24} \left(h^2-48 hH+128h-12
h_2^2+\left(12h-69\right)h_2-12h_4\right)
\end{align*}
such that $\ev'(a_i,n)=A_i(n)$ for $1\leq i\leq 5$ and $\ev'(b,n)=B(n)$. These computations lead to the following identities: For $n\geq0$,
\begin{equation}\label{Equ:A1A5BId}
\begin{aligned}
A_1(n)&=1,\quad A_2(n)=H_n,\quad A_3(n)=H_n^2,\quad
A_4(n)=\frac{1}{2}\left(H_n^{(2)}-H_n^2\right),\\
A_5(n)&=\frac{1}{2} \left(-H_n^2+2 H_n^{(2)}
   H_n-H_n\right),\\
B(n)&=\tfrac{1}{24}H_n^2-2 H_n\sum_{k=1}^n\frac{H_k}{k^2}+\tfrac{16}{3}H_n-\tfrac{1}{2}\left(H_n^{(2)}\right)^2+
\left(\tfrac{1}{2}H_n-\tfrac{69}{24}\right) H_n^{(2)}-\tfrac{1}{2}H_n^{(4)};
\end{aligned}
\end{equation}
in particular, due to Theorem~\ref{Thm:DepthOptExtIsDepthOpt}, the sum expressions on the right hands sides of~\eqref{Equ:A1A5BId} have the minimal depths $0,2,2,2,2,3$, respectively. To this end, we obtain the following identity~\cite[equ.~3.14]{Schneider:07h}: for $n\geq0$,
\begin{equation*}
\sum_{i=1}^{\infty}\frac{H_{i+n}^2}{i^2}=\frac{17}{10}\zeta_2^2
               +4H_n\zeta_3
               +H_n^2\zeta_2
               -H^{(2)}_n\zeta_2
               -\frac{1}{2}
   \left(\left(H_n^{(2)}\right)^2+H_n^{(4
   )}\right)-2H_n\sum_{k=1}^n\frac{H_k}{k^2}.
\end{equation*}

\subsection{A nontrivial harmonic sum identity}
We look for an indefinite nested sum representation of the definite sum
\begin{equation}\label{Equ:CombSum}
S(n)=\sum_{k=0}^n\binom{n}{k}^2H_k^2;
\end{equation}
for similar problems see~\cite{Schneider:06c}.
First, \SigmaP\ finds with creative telescoping the recurrence relation
\small
\begin{multline*}
   8(n+1) (2 n+1)^3 \left(64 n^4+480
   n^3+1332 n^2+1621 n+735\right)S(n)
   -4 (768 n^8+8832n^7\\
   +43056 n^6
   +115708 n^5
   +186452n^4+183201 n^3+106442 n^2+33460
   n+4533) S(n+1)\\
   +2 (n+2)(384 n^7+4224 n^6+18968 n^5+44610
   n^4+58679 n^3+42775 n^2+16084
   n+2616) S(n+2)\\
   -(n+2)(n+3)^3\left(64 n^4+224 n^3+276n^2+141n+30\right)S(n+3)=\\
   -3\left(576 n^6+4896 n^5+16660 n^4+28761
   n^3+26171 n^2+11574 n+1854\right).
\end{multline*}
\normalsize
Solving the recurrence in terms of d'Alembertian solutions and checking initial values yield the identity
$$S(n)=\binom{2n}{n}\big(\frac{1}{2}A_1(n)-\frac{19}{28}A_2(n)+B(n)\big)\quad\forall n\geq0$$
with
\begin{align*}
A_1&=\sum_{i=1}^n\frac{4 i-3}{i(2i-1)},\quad
A_2=\sum_{i=2}^n\frac{\displaystyle(4 i-3)
   \sum_{j=2}^i\frac{64 j^4-288
   j^3+468 j^2-323 j+84}{(j-1) j (2 j-3)
   (4 j-7) (4 j-3)}}{i(2 i-1)},\\
B&=-\sum_{i=2}^n\frac{\displaystyle(4 i-3)
   \sum_{j=2}^i\frac{\displaystyle\left(64
   j^4-288 j^3+468 j^2-323 j+84\right)
   \tilde{B}(j)}{(j-1) j (2
   j-3) (4 j-7) (4
   j-3)}}{i (2
   i-1)}
\end{align*}
where
$$\tilde{B}=\sum_{k=1}^j-\tfrac{3 (2 k-3) (2
   k-1) (4 k-7) \left(576 k^6-5472
   k^5+20980 k^4-41559 k^3+44882
   k^2-25113 k+5760\right)}{k \left(64
   k^4-544 k^3+1716 k^2-2379
   k+1227\right) \left(64 k^4-288 k^3+468
   k^2-323 k+84\right) \binom{2k}{k}}.$$
So far, this alternative sum representation of~\eqref{Equ:CombSum} might not be considered as really convincing. For further simplifications,
we construct the \pisiDE-field $\dfield{\QQ(x)(b)}{\sigma}$ with $\sigma(x)=x+1$ and $\sigma(b)=\frac{x+1}{2(2x+1)}b$, and we take
the
$\QQ$-monomorphism $\fct{\tau'}{\QQ(x)[b]}{\seqQ}$ from Example~\ref{Exp:Embedding3} ($\KK=\QQ$). Note that the sums $A_1,A_2,B\in\Sum_n(\QQ(x)[b])$ have the depths  $2,3,5$, respectively.
Finally, activating our machinery in this setting (we represent the sums in a \pisiDE-field and reinterpret the result by an appropriate $\QQ$-monomorphism), we arrive at the following identities: For $n\geq0$,
\begin{align*}
A_1(n)=&2\left(2 H_n-H_{2 n}\right),\\
A_2(n)=&2\left(4 H_n^2+4
   H_n+H_{2 n}^2+\left(-4 H_n-2\right)
   H_{2 n}-H_{2 n}^{(2)}\right),\\
B(n)=&\frac{3}{14}\Big(44
   H_n^2+16 H_n+11 H_{2 n}^2-\left(44
   H_n+8\right) H_{2 n}-11 H_{2
   n}^{(2)}+14
   \sum_{i=1}^n\frac{1}{i^2 \binom{2i}{i}}\Big).
\end{align*}
By Theorem~\ref{Thm:DepthOptExtIsDepthOpt} we have solved problem~\DOS\ for the expressions $A_1$, $A_2$ and $B$, i.e., we found sum representations with optimal depths $2,2,3$, respectively. Finally, this leads to the following identity: for $n\geq0$,
$$\sum_{k=0}^n\binom{n}{k}^2H_k^2=\binom{2n}{n}\left(4 H_n^2-4 H_{2
   n} H_n+H_{2 n}^2-H_{2 n}^{(2)}+
   3\sum_{i=1}^n\frac{1}{i^2\binom{2i}{i}}\right).$$


\providecommand{\bysame}{\leavevmode\hbox to3em{\hrulefill}\thinspace}
\providecommand{\MR}{\relax\ifhmode\unskip\space\fi MR }
\providecommand{\MRhref}[2]{%
  \href{http://www.ams.org/mathscinet-getitem?mr=#1}{#2}
}
\providecommand{\href}[2]{#2}

\end{document}